\documentclass[a4paper,11pt]{article}
\usepackage{graphicx} 
\usepackage{ifpdf}
   
\usepackage[all]{xy}
\usepackage{amsfonts,amsmath,amsthm,mathrsfs}
\usepackage{color}
\usepackage{subfig}
\usepackage[ansinew]{inputenc}
\usepackage[T1]{fontenc}
\usepackage{authblk}
\newtheorem{theorem}{Theorem}[section] 

\newtheorem{property}[theorem]{Property}

\title{Forest fire spreading: a nonlinear stochastic model continuous in space and time}
\author[1]{Roberto Beneduci\thanks{Email: roberto.beneduci@unical.it}}
\affil[1]{\small Dipartimento di Fisica, Universit\`a della Calabria, and INFN gruppo collegato  Cosenza, Italy}

\author[2]{Giovanni Mascali\thanks{Email: giovanni.mascali@unical.it}}
\affil[1]{\small Dipartimento di Matematica e Informatica, Universit\`a della Calabria, and INFN gruppo collegato  Cosenza, Italy}

\begin{document}
\maketitle

\begin{abstract}
Forest fire spreading is a complex phenomenon characterized by a stochastic behavior. Nowadays, the enormous quantity of georeferenced data and the availability of powerful techniques for their analysis can provide a very careful picture of forest fires opening the way to more realistic models. We propose a stochastic spreading model continuous in space and time that is able to use such data in their full power. The state of the forest fire is described by the subprobability densities of the green trees and of the trees on fire that can be estimated thanks to data coming from satellites and earth detectors. The fire dynamics is encoded into a density  probability kernel which can take into account wind conditions, land slope, spotting phenomena and so on, bringing to a system of integro-differential equations for the probability densities.  Existence and uniqueness of the solutions is proved by using Banach's fixed point theorem. The asymptotic behavior of the model is analyzed as well. 

Stochastic models based on cellular automata can be considered as particular cases of the present model from which they can be derived by space and/or time discretization. Suggesting a particular structure for the kernel, we obtain numerical simulations of the fire spreading under different conditions. For example, in the case of a forest fire evolving towards a river, the simulations show that the probability density of the trees on fire is different from zero beyond the river due to the spotting phenomenon. Firefighters interventions and weather changes can be easily introduced into the model. 

\end{abstract}










\section{Introduction}
Forest fires have a relevant role in agriculture, forestry, hydrological cycles,  soil fertility and conservation, climate stability and all of them in turn affect biological diversity. They are natural phenomena \cite{Bowman,Krebs,Walker} (disturbances) in the earth ecosystems balance and are important in the regulation of diversity and adaptive capacity of ecosystems. Moreover, forest fires behavior is influenced by the human factor, which is analyzed in several interdisciplinary studies, see Caldararo \cite{Caldararo} for a review. That brings us directly to the anthropocene crisis with humanity as a main factor for climate and ecosystems evolution. In the last decades the frequency and severity of large forest fires has increased \cite{Bowman}, and it is reasonable to expect a continuous increasing of timber emission. It has been estimated \cite{Seidl} that {\it the damages from wind, bark beetles and forest fires will increase in Europe with a rate of $0.91 \times 10^6 m^3$ of timber per year until 2030}. Yet, according to the Global Fire emission database (GFED4s), $CO_2$ emissions due to fires in 2014 were $23$ \% of global fossil fuel $CO_2$ emissions \cite{Guido}. Severe forest fires can contribute to modify the climate by emission of $CH_4$ and $N_2O$ and by emission of precursors of aerosols and ozone, moreover, they can contribute to change the albedo \cite{Bowman} and the carbon cycle \cite{Walker}. Climate change and forest fires are connected in a vicious circle since climate change make forest fires more probable, with extreme-fires becoming ever more common, while severe forest fires can speed up climate changes. All of that emphasizes the role of forest fires as a relevant factor of the anthropocene \cite{Bowman}. The impact on people and property is remarkable as well. As a consequence, forest fire management is ever more important in the effort of humanity to preserve biota and to mitigate climate change. It has been calculated \cite{Biod} that biota protection and biodiversity preservation have an enormous economic impact with annual benefits of 125-140 trillion dollars per year worldwide. In the present paper we would like to contribute to the development of mathematical modelling that constitutes one of the main tools in fire management since mathematical models can show possible fire-spread scenarios allowing planned interventions. Moreover, thanks to new technologies, mathematical models can be ever more realistic and can be confronted more easily with real scenarios. It is worth remarking that forest fires are complex phenomena which are the result of the combination of different stochastic behaviors. There exist relevant deterministic models for fire spreading, both phenomenological and physical, both based on differential equations and on cellular automata \cite{Rothermel,Richards,Finney,Tymstra,DiGregorio,Jellouli,Freire, Mendez,Denham}, see Sullivan \cite{Sullivan} for a review. Some of them are usually combined in simulation softwares. Those are good resources to figure out scenarios of the average behavior of the forest fire spreading but do not contemplate its stochastic behavior which is intrinsic to the process, particularly in its early stage when the number of firing trees is not very high and the same initial conditions can bring to different evolutions. That makes stochastic models very relevant in forest fire management as well as in epidemic management \cite{Bailey,Allen, Bertozzi,Romano} to which our model can be adapted with some appropriate modifications. In particular, a stochastic model will be able to outline different scenarios all of them compatible with the same initial conditions. 
In our knowledge, the stochastic models already developed are based on cellular automata \cite{Boychuk, Boychuk1,Stauffer,Durret,Drossel,Tinoco,Sullivan} and make some  simplifications of the real process, in particular they are discrete in space and/or time and assume that the spreading happens only in few directions that depend on the geometry of the lattice (triangular, rectangular, hexagonal, octagonal). Simplification apart, the selection of only few spreading directions could bring to distortions of the fire shape produced by the model \cite{DiGregorio,Boychuk2}.

Here, a non-linear continuous stochastic model for the spreading of forest fires in space and time is developed. The other  stochastic models based on cellular automata can be considered as particular cases that can be derived from the continuous model by space and/or time discretization. One of the advantages of the model is that it can make use in its full power of the data coming from satellites and earth detectors which are very important tools coming from new technologies and data science. Indeed, the description of the state of the fire in the model is realized by space probability density functions that can be approximated from data by means of non-parametric estimation techniques as for example kernel density estimation (KDE). Therefore, non homogeneous trees distributions can be taken into account making the model and its predictions closer to the real features of the forest. The dynamics is encoded into a density probability kernel which takes into account both wind conditions, land slope, spotting phenomena, and brings to a system of integro-differential equations for the probability densities whose time evolution can be forecasted once the system has been solved. The numerical implementation of the model provides probabilistic space-time maps of the fire evolution. Since the dynamics of the forest fire can be detected by means of data acquisition and elaboration through KDE, a very accurate model validation is at hand.
Besides, the model we propose can be used to outline possible scenarios compatible with the initial conditions. Moreover, the model is able to describe fire spotting which is a typical stochastic mechanism for fire spreading at long distances and to include firefighters interventions and weather changes in its dynamics. It is worth remarking that, at variance with cellular automata models, fire spreading is not constrained in few directions, but is modelled as a continuous process in all directions giving a more realistic description of fire spreading. One of the motivation for using cellular automata is that the simplification they make in describing fire spreading corresponds to a reduction in the computational costs. Here we show that the computational problem one has to face when dealing with a continuous model can be overcome due to the possible parallelization of the numerical scheme. Moreover, by Banach's fixed point theorem, it is proved that the non-linear system of integro-differential equations that encodes the dynamics of the forest fires admits a unique solution. That is relevant from the mathematical viewpoint and for the consistency of the model. In the next section we build the model and propose a physical interpretation. Then, we provide its numerical implementation (by using fourth order Runge-Kutta scheme) and show some solutions in simple scenarios. In the appendix we prove existence and uniqueness of solutions and analyze the asymptotic behavior.

\section{The Probabilistic Model}
\label{themodel}

\begin{figure}[tbhp]
\centering
\includegraphics[scale=0.7]{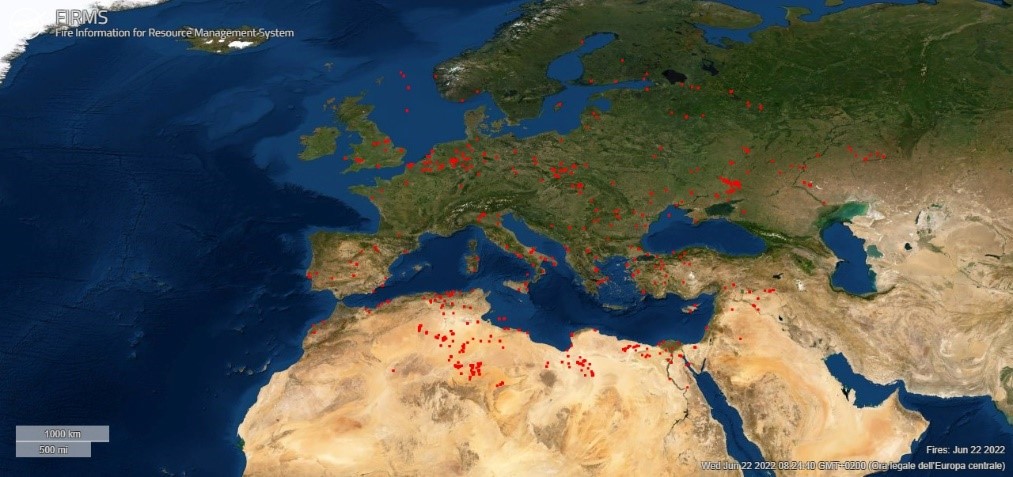}
\includegraphics[scale=0.7]{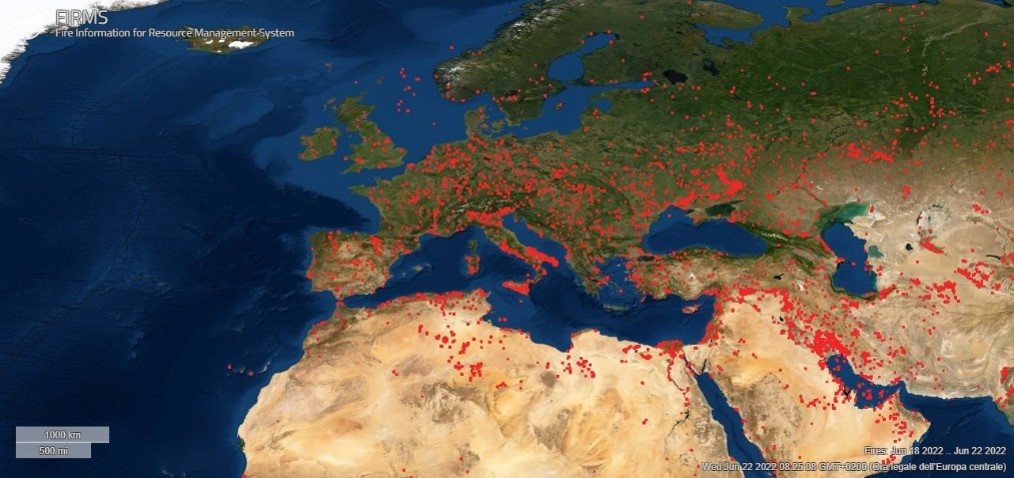}
\caption{Qualitative illustration of the spatio-temporal evolution of a forest fire.  The image illustrates qualitatively the expansion of the forest fire in Europe during the period 18-22 June 2022 and can be interpreted as a probability cloud evolving in time. The image has been downloaded from the NASA website {https://shorturl.at/pKTV5}.}
\label{fig1}
\end{figure}

 Let $\Sigma$ be the surface of the earth we focus on. A point in the surface is denoted by $\mathbf{x}=(x,y)\in\Sigma$. We assume there are positive definite functions $\psi^F(t,\mathbf{x}),\psi^B(t,\mathbf{x}),\psi^G(t,\mathbf{x})\in L^1(\Sigma,d^2\mathbf{x})$ representing sub-probability densities and describing the state of the forest fire at time $t$. In particular, $\psi^F(t,\mathbf{x})$ is the sub-probability density for the trees on fire, $\psi^B(t,\mathbf{x})$ the sub-probability density for the burnt trees (firing trees which turn into burnt ones) and $\psi^G(t,\mathbf{x})$ the sub-probability density for the green trees.  The probability density functions are assumed to be derivable with respect to the time variable. Moreover, we assume $\psi^G(0,\mathbf{x})+\psi^F(0,\mathbf{x})=\psi^F(t,\mathbf{x})+\psi^G(t,\mathbf{x})+\psi^B(t,\mathbf{x})$ for every $t$ and $\int_\Sigma(\psi^G(0,\mathbf{x})+\psi^F(0,\mathbf{x}))\,d\mathbf{x}=1$. It is worth noticing that the existence of $\psi^F$, $\psi^B$ and $\psi^G$ is a theoretical assumption; the actual fire distribution is just an instantiation of the random process associated to the fire distribution, see Fig. \ref{fig1} which illustrates the spatio-temporal evolution of a forest fire. Anyway, an approximation of $\psi^F$, $\psi^B$ and $\psi^G$ can be obtained through density estimation methods as for example the kernel density estimation (KDE) \cite{Givens} which provides differentiable probability density functions, starting from a distribution like that in Fig. \ref{fig1}.

The model is inspired by an epidemic model \cite{B} where an interpretation of the probability densities motivated by the statistical interpretation of quantum mechanics \cite{Bal70} is suggested. We remark that in the present paper the focus is on the probability density functions $\psi^F$, $\psi^G$, $\psi^B$ which, by analogy with quantum mechanics, define the state of the forest fire. We assume the existence of these probability density functions for which we postulate a deterministic evolution law (see below).

Moreover, the model,  describing the evolution of $\psi^F$ can be used to locate those regions of space where the probability of firing is going to increase or to decrease. That could be very helpful to governments as we  remarked.

Now, we proceed to define the evolution equation for the state of the forest fire. We provide a  non-linear evolution for the sub-probability distributions. 

\noindent
The time it takes for a firing tree to burn down depends on several factors (humidity, nature of the wood, wind exposure, distance to other trees, mass, mass distribution, etc.) that make it a random variable, which we indicate by $T$. Let $p(t)$ denote its probability distribution that we assume to be continuous. Therefore, the probability that a tree that started firing at $t=0$ burns down in the time interval $[0,t]$ is $P(T<t)=F(t)=\int_0^t p(\tau)\,d\tau$ and the probability that the same tree is still firing at time $t$ is $\pi(t)=F(\infty)-F(t)=P(T>t)$. Then, by the Bayes theorem, the probability that a tree burns down during the time interval $[t,t']$, $t'>t$, given it is firing at time $t$, is $P(t<T<t'\vert T>t)=\frac{F(t')-F(t)}{F(\infty)-F(t)}$. We can define the burning rate, that is, the conditional probability rate $\alpha(t):=\lim_{\epsilon\to 0}\frac{P(t<T<t+\epsilon\vert T>t)}{\epsilon}=\frac{1}{F(\infty)-F(t)}\lim_{\epsilon\to 0}\frac{F(t+\epsilon)-F(t)}{\epsilon}=\frac{1}{\pi(t)}\frac{dF}{dt}=\frac{p(t)}{\pi(t)}$.

\noindent
In the case the probability distribution of T is exponential, $p(t)=\lambda e^{-\lambda t}$, we obtain $\alpha(t)=\lambda$ and the average time it takes for a firing tree to burn down is $\hat{F}=\frac{1}{\lambda}$, 
where the parameter $\lambda$ can therefore depend on the factors specified above. Moreover, in the exponential case, the burning down process is memoryless; in other words, the probability  that a tree that is firing at time t will fire down in the time interval $[t,t']$ does not depend on $t$ but only on $t'-t$.

\subsection{Modelling the burning down process}
The probability rate $\alpha(t)$ can be used to model the burning down process and to connect $\psi^B$ and $\psi^F$: 

\begin{equation}\label{alpha}
	\psi^B(t,\mathbf{x})=\int_0^t\alpha(\tau)\psi^F(\tau,\mathbf{x})d\tau,
\end{equation}
\noindent 
we notice that $\alpha$ can also be a function of the position, since the trees in a location can be different from the trees in another location (e.g., type of wood, humidity and so on).

\noindent
Note that the time $\tau$ in equation (\ref{alpha}) is the forest fire time. If we assume that the burning down process of a tree does not depend on the time the tree started burning, $\alpha$ must be a constant (memoryless process). On the contrary, in order to describe a scenario in which external factors can influence the forest fire propagation, the time dependence of $\alpha$ should be introduced. For example, firefighters intervention will certainly increase the probability that a tree stops firing; that would correspond to an increasing function $\alpha(t)$. Later (see the subsection below regarding the model with memory) we will modify the $\psi^B$ in order to describe a scenario with no external intervention but with a burning down process which is not memoryless.

 \subsection{Modelling the fire spreading}
The forest fire propagation is encoded into a kernel $W(t,\mathbf{x},\mathbf{y})$ which describes the transition probability of the fire from $\mathbf{y}$ to $\mathbf{x}$ at time $t$. We remark that the transition probability takes into account the randomness of the forest fire; just to give some examples: 1) it considers the randomness of the blaze geometry, 2) firebrand spotting in which airborne burning particles, which travel through wind-driven transport, can ignite new fires beyond the fire line, 3) the randomness in the heat propagation due to its turbulence.   

In particular, the quantity   

$$[d^2\mathbf{x}\psi^G(\tau,\mathbf{x})W(\tau,\mathbf{x},\mathbf{y})d\tau]$$

\noindent
is interpreted as the conditional probability that a green tree in the neighbourhood centered in $\mathbf{x}$ with radius $d^2\mathbf{x}$  is set on fire during the time interval $[\tau,\tau+d\tau]$ if a tree on fire is present in $\mathbf{y}$ at time $\tau$.{\footnote{The meaning of $W$ can be better understood by introducing the quantity 
$$\mu_{\Delta_2}(\tau,\mathbf{y})\,d\tau:=\int_{\Delta_2}[W(\tau,\mathbf{x},\mathbf{y})d\tau]\psi^G(\tau,\mathbf{x})d^2\mathbf{x}$$
\noindent
which can be interpreted as the transition probability that a green tree in spatial region $\Delta_2$ is set on fire during the time interval  $(\tau,\tau+d\tau)$ if in $\mathbf{y}$ there is a tree on fire. 
Indeed, $W(\tau,\mathbf{x},\mathbf{y})$ is the Radon-Nikodym derivative of $\mu_{(\cdot)}(\tau,\mathbf{y})$ with respect to  $\int_{(\cdot)}\psi^G(\tau,\mathbf{x})\,d^2\mathbf{x}$;
$\mu_{(\cdot)}(\tau,\mathbf{y})$ being  absolutely continuous with respect to $\int_{(\cdot)}\psi^G(\tau,\mathbf{x})\,d^2\mathbf{x}$. That seems quite natural since $\int_{\Delta_2}\psi^G(\tau,\mathbf{x})\,d^2\mathbf{x}=0$ implies that no green trees can be set on fire inside $\Delta_2$. }
Then, 
\begin{equation}\label{transition}
[d^2\mathbf{x}\psi^G(\tau,\mathbf{x})W(\tau,\mathbf{x},\mathbf{y})d\tau](\psi^F(\tau,\mathbf{y})\,d^2\mathbf{y})
\end{equation}

\noindent
represents the probability that a green tree in the neighbourhood centered in $\mathbf{x}$ with radius $d^2\mathbf{x}$  is set on fire by the trees on fire contained in the neighbourhood centered in $\mathbf{y}$ with radius $d^2\mathbf{y}$ during the time interval $[\tau,\tau+d\tau]$.  Hence, 

\begin{equation}\label{t}
d^2{\mathbf{x}}\int_{\Sigma}\int_0^{t}\,\psi^G(\tau,\mathbf{x})W(\tau,\mathbf{x},\mathbf{y})\psi^F(\tau,\mathbf{y})\,d\tau\,d^2\mathbf{y}
\end{equation}

\noindent 
gives the probability of new trees on fire in the neighbourhood centered in $\mathbf{x}$ with radius $d^2\mathbf{x}$ during the time interval $[0,t]$. 

Since in the same time interval there is a probability that the trees fire down, we have

 \begin{align}\label{bilancio}
d^2{\mathbf{x}}\int_{\Sigma}\int_0^{t}\,\psi^G(\tau,\mathbf{x})&W(\tau,\mathbf{x},\mathbf{y})\psi^F(\tau,\mathbf{y})\,d\tau\,d^2\mathbf{y}=\notag\\
&=d^2{\mathbf{x}}\,[\psi^F(t,\mathbf{x})-\psi^F(0,\mathbf{x})+\psi^B(t,\mathbf{x})].
\end{align}

\noindent

Concerning the time dependence of $W$, it can be due, for example, to changes in wind and atmospheric conditions, or to external interventions such as firefighters action which will certainly decrease the probability that a firing tree can make a green tree to start firing; that would correspond to a decreasing function $W(t)$.

The flux diagram in figure \ref{flux}, case 1), illustrates the variation of the sub-probability distributions during an infinitesimal time interval $dt$.

\begin{figure}[ht]
\centering
\includegraphics[scale=0.3]{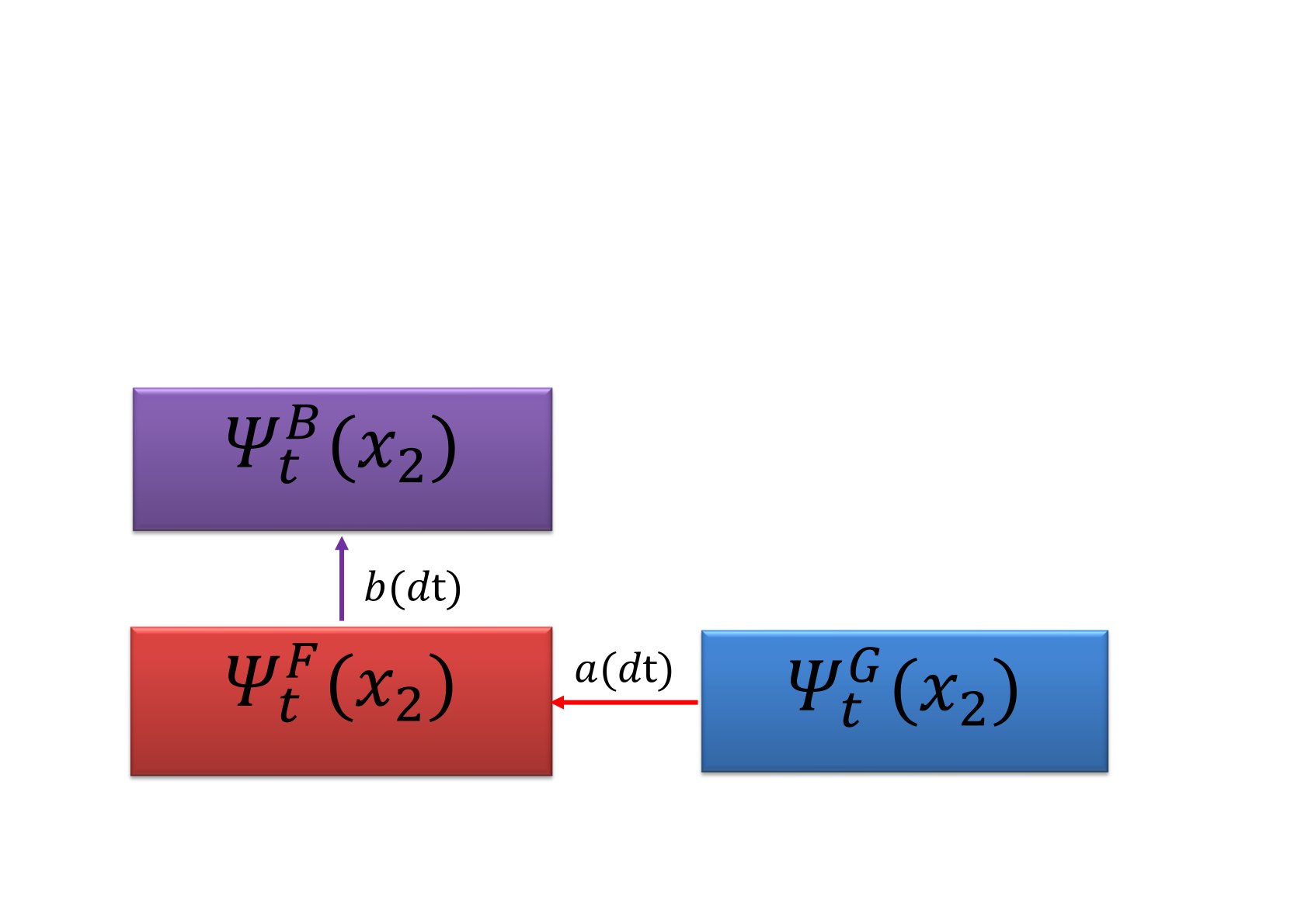}
\caption{Evolution of the sub-probability densities during an infinitesimal time interval $d\,t$. The arrows represent the direction of the probability fluxes. \\
1) Memoryless model: $a\,dt=\int_{\Sigma}\psi^G(t,\mathbf{x})W(t,\mathbf{x},\mathbf{y})\psi^F(t,\mathbf{y})dt\,d^2\mathbf{y}$ and $b\,dt=\alpha(t,\mathbf x)\psi^F(t,\mathbf x)dt$.\\
2) Model with memory: $a\,d\,t=\int_{\Sigma}\psi^G(t,\mathbf{x})W(t,\mathbf{x},\mathbf{y})\psi^F(t,\mathbf{y})d^2\mathbf{y}\,dt$ and $b\,dt=-\dot\pi(t)\psi^F(0,\mathbf{x})dt
$ $-\Bigg[\int_0^t\int_{\Sigma}\psi^G(\tau,\mathbf{x})W(\tau,\mathbf{x},\mathbf{y})$ $\times\psi^F(\tau,\mathbf{y}) \dot\pi(t-\tau)d\tau\,d^2\mathbf{y}\Bigg]dt$.}
\label{flux}
\end{figure}

\vskip1cm

\noindent  
 By (\ref{bilancio}) we obtain,

 \begin{align}\label{evolution}
 \psi^F(t,\mathbf{x})&=\psi^F(0,\mathbf{x})-\psi^B(t,\mathbf{x})+\int_{\Sigma}\int_0^{t}\,\psi^G(\tau,\mathbf{x})W(\tau,\mathbf{x},\mathbf{y})\psi^F(\tau,\mathbf{y})\,d\tau\,d^2\mathbf{y}
\end{align}

\noindent
with,

\begin{align}\label{balance}
\psi^G(0,\mathbf{x})+\psi^F(0,\mathbf{x})&=\psi^F(t,\mathbf{x})+\psi^G(t,\mathbf{x})+\psi^B(t,\mathbf{x})\\
1&=\int_\Sigma\big(\psi^G(0,\mathbf{x})+\psi^F(0,\mathbf{x})\big)\,d^2\mathbf{x}\notag\\
\psi^B(t,\mathbf{x})&=\int_0^t\alpha(\tau,\mathbf x)\psi^F(\tau,\mathbf{x})\,d\tau.\notag
\end{align}

\noindent
By (\ref{balance}) and (\ref{evolution}) we obtain,

\begin{align}\label{evolutionG}
	\psi^G(t,\mathbf{x})=\psi^G(0,\mathbf{x})-\int_{\Sigma}\int_0^{t}\,\psi^G(\tau,\mathbf{x})W(\tau,\mathbf{x},\mathbf{y})\psi^F(\tau,\mathbf{y})\,d\tau\,d^2\mathbf{y}.
\end{align}

\noindent
Equations (\ref{evolution}), (\ref{evolutionG}) and (\ref{alpha}) define the following non-linear system of integral equations

\begin{equation}\label{system}
	\begin{cases}
		\psi^F(t,\mathbf{x})&=\psi^F(0,\mathbf{x})-\psi^B(t,\mathbf{x})+\int_{\Sigma}\int_0^{t}\,\psi^G(\tau,\mathbf{x})W(\tau,\mathbf{x},\mathbf{y})\vert\psi^F(\tau,\mathbf{y})\vert\,d\tau\,d^2\mathbf{y}\\
		\psi^G(t,\mathbf{x})&=\psi^G(0,\mathbf{x})-\int_{\Sigma}\int_0^{t}\,\psi^G(\tau,\mathbf{x})W(\tau,\mathbf{x},\mathbf{y})\vert\psi^F(\tau,\mathbf{y})\vert\,d\tau\,d^2\mathbf{y}.
	\end{cases}
\end{equation}
See the appendix for the proof of the existence and uniqueness of solutions of system (\ref{system}).
\noindent

System (\ref{system}) can be written in differential form as follows. Supposing $\epsilon$ sufficiently small, we can use (\ref{evolution}) to derive the evolution from the state at time $t$ to the state at time $t+\epsilon$,

\begin{align}\label{step}
\psi^F(t+\epsilon,\mathbf{x})&=\psi^F(0,\mathbf x)-\int_0^{t+\epsilon}\alpha(\tau)\psi^F(\tau,\mathbf x)\,d\tau+ \int_{\Sigma}\int_0^{t+\epsilon}\,\psi^G(\tau,\mathbf{x})\,W(\tau,\mathbf{x},\mathbf{y})\psi^F(\tau,\mathbf{y})\,d\tau\,d^2\mathbf{y}\notag\\
&=\psi^F(0,\mathbf x)-\int_0^{t}\alpha(\tau)\psi^F(\tau,\mathbf x)\,d\tau-\int_t^{t+\epsilon}\alpha(\tau)\psi^F(\tau,\mathbf x)\,d\tau+\nonumber\\
&\quad\quad\quad\quad\quad+\int_{\Sigma}\int_0^{t}\,\psi^G(\tau,\mathbf{x})\,W(\tau,\mathbf{x},\mathbf{y})\psi^F(\tau,\mathbf{y})\,d\tau\,d^2\mathbf{y}+\notag\\
&\quad\quad\quad\quad\quad+\int_{\Sigma}\int_t^{t+\epsilon}\,\psi^G(\tau,\mathbf{x})\,W(\tau,\mathbf{x},\mathbf{y})\psi^F(\tau,\mathbf{y})\,d\tau\,d^2\mathbf{y}\notag\\
&=\psi^F(t,\mathbf{x})-\alpha(t,\mathbf{x})\psi^F(t,\mathbf{x})\,\epsilon+\epsilon \psi^G(t,\mathbf{x})\int_{\Sigma}\,W(t,\mathbf{x},\mathbf{y})\psi^F(t,\mathbf{y})\,d^2\mathbf{y}.
\end{align}  

\noindent
Then, dividing both sides by $\epsilon$ and taking the limit for $\epsilon$ going to $0$, one gets

\begin{equation}\label{intdiff}
\frac{\partial\psi^F(t,\mathbf{x})}{\partial t}=-\alpha(t,\mathbf{x})\psi^F(t,\mathbf{x})+\psi^G(t,\mathbf{x})\int_{\Sigma}\,W(t,\mathbf{x},\mathbf{y})\psi^F(t,\mathbf{y})\,d^2\mathbf{y}.
\end{equation}

\noindent
Using a similar procedure, one can obtain the differential equation for the green trees probability density

\begin{equation}\label{intdiff1}
\frac{\partial \psi^G_{t}(\mathbf{x})}{\partial t}=-\psi^G_{t}(\mathbf{x})\int_{\Sigma}W(t,\mathbf{x},\mathbf{y})\psi^F(t,\mathbf{y})d^2\mathbf{y}.
\end{equation}
\noindent

\noindent
Equations (\ref{intdiff}) and (\ref{intdiff1}) define a non linear system of integro-differential equations.

\subsection{The case with memory}

\noindent
In the case one needs to describe a burning down process whose probability rate depends on the time (memory), it is possible to modify the model by changing the way we calculate $\psi^B$. In particular, recalling that $\pi(t)$ is the probability that a tree that went on fire at time $t=0$ is still on fire at time $t$, the probability that the trees that went on fire at position $\mathbf x$ during the time interval $[\tau,\tau+d\tau]$ have not completely burned at time $t>\tau$ is given by

\begin{equation}\label{t1}
\int_{\Sigma}\,\Big[\psi^G(\tau,\mathbf{x})W(\tau,\mathbf{x},\mathbf{y})\psi^F(\tau,\mathbf{y})\Big]\pi(t-\tau)\,d\tau.
\end{equation}

\noindent
As a consequence, the flux diagram has to be modified, see figure \ref{flux} case 2).

On the basis of these considerations we can write down the evolution equations for the sub-probability densities $\psi^F$ and $\psi^G$, we have
\begin{align}
\label{evolutionF}
\psi^F(t,\mathbf{x})&=\psi^F(0,\mathbf{x})\pi(t)+\int_0^{t}\,\int_{\Sigma}\psi^G(\tau,\mathbf{x})W(\tau,\mathbf{x},\mathbf{y})
\psi^F(\tau,\mathbf{y})\pi(t-\tau)\,d\tau\,d^2\mathbf{y},\\
\psi^G(t,\mathbf{x})&=\psi^G(0,\mathbf{x})-\int_0^t\int_\Sigma\psi^G(\tau,\mathbf{x})W(\tau,\mathbf{x},\mathbf{y})\psi^F(\tau,\mathbf{y})\,d\tau\,d^2\mathbf{y}.
\label{evolutionG1}
\end{align}
\noindent
The sub-probability density $\psi^B$ of burned trees is defined by
\begin{align}
\psi^G(0,\mathbf{x})+\psi^F(0,\mathbf{x})=:\psi^G(t,\mathbf{x})+\psi^F(t,\mathbf{x})+\psi^B(t,\mathbf{x}),
\end{align}
and the following property holds
\begin{align*}
\int_\Sigma\big(\psi^G(0,\mathbf{x})+\psi^F(0,\mathbf{x}))\,d^2\mathbf{x}=\int_\Sigma\big(\psi^G(t,\mathbf{x})+\psi^F(t,\mathbf{x})+\psi^B(t,\mathbf{x})\big)\,d^2\mathbf{x}=1.
\end{align*}

\noindent
Equations (\ref{evolutionF}) and (\ref{evolutionG1}) constitute a non-linear system of integral equations.

By the same procedure we used to derive equations (\ref{intdiff}) and (\ref{intdiff1}), the system of equations  (\ref{evolutionF})-(\ref{evolutionG1}) can be written in differential form. Supposing $\epsilon$ sufficiently small, we can derive the evolution from the state at time $t$ to the state at time $t+\epsilon$,
\begin{align*}
\psi^F(t+\epsilon,\mathbf{x})&=\psi^F(0,\mathbf{x})\pi(t+\epsilon)+
\int_0^{t+\epsilon}\int_{\Sigma}\psi^G(\tau,\mathbf{x})\,W(\tau,\mathbf{x},\mathbf{y})\psi^F(\tau,\mathbf{y})\,\pi(t+\epsilon-\tau)\,d\tau\,d^2\mathbf{y}\\
&\approx\psi^F(0,\mathbf{x})[\pi(t)+\epsilon\dot\pi(t)]+
\int_{\Sigma}\int_0^{t}\psi^G(\tau,\mathbf{x})\,W(\tau,\mathbf{x},\mathbf{y})\psi^F(\tau,\mathbf{y})\,[\pi(t-\tau)+\epsilon\dot\pi(t-\tau)]\,d\tau\,d^2\mathbf{y}\\
&+\int_t^{t+\epsilon}\int_{\Sigma}\psi^G(\tau,\mathbf{x})\,W(\tau,\mathbf{x},\mathbf{y})\psi^F(\tau,\mathbf{y})\,\pi(t-\tau)\,d\tau\,d^2\mathbf{y}.\\
\end{align*}  

\noindent
Then, we can get an integro-differential equation for the firing trees probability density $\psi^F$. Indeed $\psi^F(t+\epsilon,\mathbf{x})=\psi^F(t,\mathbf{x})+\epsilon\frac{\partial\psi^F(t,\mathbf{x})}{\partial t}+o(\epsilon)$ so that substituting the latter in the previous expression,
dividing both sides by $\epsilon$ and taking the limit for $\epsilon \rightarrow 0$, we obtain
\begin{eqnarray}
\nonumber
\frac{\partial\psi^F(t,\mathbf{x})}{\partial t}&=&\psi^F(0,\mathbf{x})\dot\pi(t)+
\int_{\Sigma}\int_0^{t}\psi^G(\tau,\mathbf{x})\,W(\tau,\mathbf{x},\mathbf{y})\psi^F(\tau,\mathbf{y})\,\dot\pi(t-\tau)\,d\tau\,d^2\mathbf{y}\\
&+&\int_{\Sigma}\psi^G(t,\mathbf{x})\,W(t,\mathbf{x},\mathbf{y})\psi^F(t,\mathbf{y})\,d^2\mathbf{y}. \label{intdiffF}
\end{eqnarray}

Analogously, it is possible to write an integro-differential equation for the green trees probability density $\psi^G$,

\begin{eqnarray}
\frac{\partial\psi^G(t,\mathbf{x})}{\partial t}&=&-
\int_{\Sigma}\psi^G(t,\mathbf{x})\,W(t,\mathbf{x},\mathbf{y})\psi^F(t,\mathbf{y})\,d^2\mathbf{y}. \label{intdiffG}
\end{eqnarray}
For the proof of the existence and uniqueness of solutions of the previous system, see Appendix.

In the case of a memoryless burning process, $\pi(t)=e^{-\lambda t}$, and it can  be easily seen that equations (\ref{intdiffF}) and (\ref{intdiffG}) coincide with equations (\ref{intdiff}) and (\ref{intdiff1}) respectively.

\section{Numerical simulations}
\label{numsim}
In this section we will investigate the potentialities of the proposed model by presenting the results of some numerical simulations.
An essential element of the model is the rate function $W$ which has to incorporate
the main factors which influence fire spread, such as the type of vegetation, terrain slope, humidity, wind speed, as well as spotting phenomenon.
It is not easy to take into account all these factors, since they are not completely understood and, furthermore, they interact with each other. 
We attempt at a phenomenological approach inspired by some theoretical results \cite{Sardoy,Perry}. 
In particular, we assume that $W$ can be written as the sum of two terms: $W_1(t,\mathbf x,\mathbf y)$ and $W_2(t,\mathbf x,\mathbf y)$.
The former one takes into account the fire transition probability rate from $\mathbf y$ to $\mathbf x$ due to heat transfer by radiation or convection and can be written as
$$
W_1(t,\mathbf x,\mathbf y)=c_{w_1}\exp{[-\rho^2/2 \sigma_{w_1}^2(1+e\cos{(\varphi-\bar\varphi)})]}.
$$
where $\rho:=|\mathbf x-\mathbf y|$, $c_{w_1}$ and $ \sigma_{w_1}$ can depend on the space and time variables, $e$ is a sort of eccentricity taking into account
that fire spreads mainly in the direction of the resultant wind-slope vector $\mathbf v$, 
while $\bar \varphi$ and $\varphi$ respectively are the angles that the vectors $\mathbf v$ and $\mathbf x-\mathbf y$ makes with the $x$-axis. The vector $\mathbf v$ is given by
$$
\mathbf{v}=\mathbf{v}_w+\mathbf{v}_s,
$$ 
where $\mathbf{v}_w$ is the component in the
wind direction, whose modulus quantifies the wind effect, while $\mathbf{v}_s$  is in the direction of the maximum slope at $\mathbf y$ and has modulus equal to the slope effect. Wind and slope effects are respectively
given by:
\begin{eqnarray*}
&&	|\mathbf{v}_w|=C(3.281 \nu)^B(\beta/\beta_{op}^{-E}),\\
&&  |\mathbf{v}_s|=5.275 \beta^{-0.3} \tan \omega^2,
\end{eqnarray*}
here $\nu$ is the wind speed, $\beta$ is the packing ratio of the fuel bed, $\beta_{op}$ is the optimum packing ratio, $\omega$ 
is the slope at $\mathbf y$ expressed in radians, and $C$, $B$ and $E$ are coefficients which depend on the fuel particle size in the fuel bed \cite{Tru04}.

The functions $c_{w_1}$, $\sigma_{w_1}$ and the eccentricity $e$ depend on the above-specified factors. This dependence might be accounted for into physical models or, empirically, by the fitting of real data. Here, we do not enter into the details and limit ourselves to develop the spreading model and its interpretation; its implementation in real scenarios, for which a deeper description of the parameters is necessary, will be the topic of a future work.

Concerning the function $W_2$, it represents the fire transition probability rate from $\mathbf y$ to $\mathbf x$ due to firebrand generation, transport, and ignition. We assume that firebrand generation is a Poisson process \cite{Gnedenko} with density\footnote{In a Poisson process, the density is constant. Here, the time dependence in $c_{w_2}(\mathbf{y},t)$ is introduced to take account of possible variations in the emission rate due to variations in the atmospheric conditions and/or external interventions (which would correspond to a Poisson process with a different density) while the space dependence can be due to the fact that different space regions can contain different types of trees.} $c_{w_2}(\mathbf{y},t)$. Therefore, the emission probability during a sufficiently small time interval $[t,t+\Delta t]$ does not depend on $t$ but only on $\Delta t$. In other words, we model a tree as a Poissonian emitter. That is an idealization since the emission probability should depend on the firing intensity of the tree and then on the time it started firing. Anyway such an idealization should be effective in describing the fire spotting process since we focus on the average behavior of the trees.  According to the Poisson process theory, the probability that a single firebrand is emitted during the time interval $[t,t+\Delta t]$, given that it is firing at time $t$, is $P(Emission\,\,in\,\,[t,t+\Delta t]|Firing\,\, at\,\, \mathbf{y}):=c_{w_2}(\mathbf{y},t)\Delta t+o(\Delta t)$, while the probability that more than one firebrand is emitted is proportional to $o(\Delta t)$. It follows that the probability that a tree in a neighborhood of $\mathbf{y}$ is firing at time $t$ and emits a firebrand during the time interval $[t,t+\Delta t]$ is

$$P(emission\,\,in\,\,[t,t+\Delta t]|firing\,\,at\,\,\mathbf{y})\psi^F(t,\mathbf{y})\,d\mathbf y=c_{w_2}(\mathbf{y},t)\psi^F(t,\mathbf{y})\Delta t\,d\mathbf y.$$

 \noindent
Further, we assume $W_2$ to have the form
\begin{eqnarray}
W_2=
\left\{
\begin{array}{l}
\frac{c_{w_2}}{\rho}\exp{\Big[- \frac{(\log{(\rho/\rho_0)}-\mu )^2}{2 \sigma_{w_2}^2} \Big] }\exp{\Big(-\frac{1}{\cos{(\varphi-\bar\varphi)}}\Big)} 
\exp{\Big(-\Big(\frac{((\mathbf x-\mathbf y)\cdot \frac{\mathbf v_w}{|\mathbf v_w|})^2}{2\sigma_1^2}+\frac{(\mathbf x-\mathbf y)\cdot {\mathbf n_w}}{2\sigma_2^2}\Big)\Big)},
\,\, -\frac{\pi}{2}<\varphi< \frac{\pi}{2}\\
0,\qquad \textrm{otherwise},
\end{array}
\right.
\label{W2}
\end{eqnarray}
where 
$	\frac{1}{\rho}\exp{\Big[- \frac{(\log{(\rho/\rho_0)}-\mu )^2}{2 \sigma_{w_2}^2} \Big] }$
$\times\exp{\Big(-\frac{1}{\cos{(\varphi-\bar\varphi)}}\Big)}$ is proportional to the probability that a firebrand generated at $\mathbf y$ lands at $\mathbf x$ \cite{Sardoy,Perry}, in particular the first two factors represent a lognormal function of the landing distance \cite{Sardoy,Men} while the third factor takes into account that this probability has its maximum in the wind direction and approaches zero as the angle between $\mathbf x -\mathbf y$ and this direction goes to $\frac{\pi}{2}$. The last factor is proportional to the probability of fire ignition at $\mathbf x$, given that in $\mathbf{x}$ there is a green tree,

$$\exp{\Big(-\Big(\frac{((\mathbf x-\mathbf y)\cdot \frac{\mathbf v_w}{|\mathbf v_w|})^2}{2\sigma_1^2}+\frac{(\mathbf x-\mathbf y)\cdot {\mathbf n_w}}{2\sigma_2^2}\Big)\Big)}\psi^G(t,\mathbf{x})d^2\mathbf{x}$$

\noindent
being the probability that a tree in a neighborhood of $\mathbf{x}$ is put on fire by a firebrand coming from $\mathbf{y}$. This probability exponentially decreases with $\rho$,
since the longest the travelled distance the greater is the decreasing in the temperature of the firebrands. It is also taken into account that the velocity in the wind direction is higher, hence the same distance is travelled in less time and the decreasing of the temperature is lower.
 
Therefore, combining all those conditional probabilities, we obtain that

$$dt\,d^2\mathbf{x}\int_\Sigma \psi^G(t,\mathbf{x})W_2\psi^F(t,\mathbf{y})d^2\mathbf{y}$$

\noindent
can be interpreted as the probability that a firebrand arrives in a neighbourhood of $\mathbf{x}$ and makes a green tree to start firing during the time interval $[t,t+dt]$.

The meaning of the various terms which appear in \eqref{W2} is the following
\begin{itemize}
\item $\rho_0$ is a reference length, $\mu$ and $\sigma_{w_2}$ are the mean and the standard deviation of $\log{\frac{\rho}{\rho_0}}$ respectively;
\item $\mathbf n_w$ is the unit vector in the direction orthogonal to that of the wind, and $\sigma_1$ and $\sigma_2$ are the standard deviations
of the probability of fire ignition in the wind direction and in the direction orthogonal to it.
\end{itemize}
The terms $c_{w_2}$ $\mu$, $\sigma_{w_2}$, $\sigma_1$ and $\sigma_2$ depend on the various factors which influence the firebrand phenomenon such as 
the critical surface fire intensity needed to ignite the crown of a tree, the surface fireline intensity, the wind speed, the fire intensity
produced by torching trees and so on. Also this dependence will be the object of future studies.

We conducted two preliminary numerical experiments. In the first one we considered only the function $W_1$, while in the second one we took into account also $W_2$.
For simplicity, in both experiments we considered a flat rectangular wildland and a constant wind speed, which points along the negative $x$-axis. Moreover, $c_{w_1}$, $\sigma_{w_1}$, $e$, $c_{w_2}$,$\sigma_{w_2}$, $\sigma_1$,
and $\sigma_2$ were taken to be constant. The unit time (u.t.) was taken equal to the decay time of the firing trees, while the unit length (u.l.) was chosen in such a way that a 1$\times$1 square in the spatial grid corresponds to the mean area of a tree which is around 15m$^2$. The values of the constants are reported in Table \ref{tab:1} below.

\begin{table}[h!]
\caption{Values of the terms present in the expressions of $W_1$ and $W_2$. The terms without units are dimensionless.}
\label{tab:1}       
\begin{center}
	\begin{tabular}{|c | c || c | c|| c |c || c | c |}
		\hline  \noalign{\smallskip}
		$c_{w_1}$& 1$\times$10$^3\,\,$1/u.t. & $\bar\varphi$  &$\pi$ rad  & $\mu$ &  25 & $\sigma_2$& 4.86 $\times$ 10$^{-2}$\,\, u.l.  \\
		$\sigma_{w_1}$ &1.41 $\times$ 10$^{-2}$\,\, u.l. & $c_{w_2}$ &  1$\times$10$^3\,\,$u.l./u.t.  & $\sigma_{w_2}$ & 5& & \\
		$e$ & 0.9& $\rho_0$ & 2.38 $\times$ 10$^{-4}$\,\, u.l.   & $\sigma_1$ & 0.391\,\, u.l.   &  &\\
		\noalign{\smallskip}\hline
	\end{tabular}
\end{center}
\end{table}.

At the middle of the rectangular land there is a subregion, parallel to the $y$-axis, which is not inflammable, for example a river. In both cases we took $\pi(t)=\exp{(-t)}$. The initial green trees sub-probability density is constant in the inflammable region, while that for the firing trees is non-null only in a small area centered at $x_1=1.92\, u.l.$, $x_2=0.6\, u.l.$, see Fig. \ref{fig_i_1} below. From Figs. \ref{fig_i_1}-\ref{fig_i_6} below, it can be seen that if the spotting phenomenon is not taken into account, the fire propagates upwind and the wildfire growth is ovoid. The fire reaches the non-inflammable region but does not manage to overcome it, and slowly propagates downwind until it is extinguished. From Fig. \ref{fig_i_6} one can see that there is a non-null probability that a part of the trees survives and that this probability increases near the right boundary of the region. When the term $W_2$ is added to the transition rate the fire manages to jump the non-inflammable region, see Figs. \ref{fig_i_7}-\ref{fig_i_12} below.
We numerically solved the system of equations (\ref{intdiffF})-(\ref{intdiffG}), by using a fourth order Runge-Kutta scheme. We used a spatial grid consisting of 251$\times$ 121
points and a time step equal to 10$^{-2} u.t.$

\noindent
Eventually, we remark that the numerical simulation could be very much sped-up by exploiting the inherent parallelism of the model, in fact the spatial integrals which appear on the right-hand sides of equations (\ref{intdiffF})-(\ref{intdiffG})  can be computed at any point of the grid independently from each other.

{\begin{figure}
\begin{center}
	\includegraphics[width=0.45\textwidth]{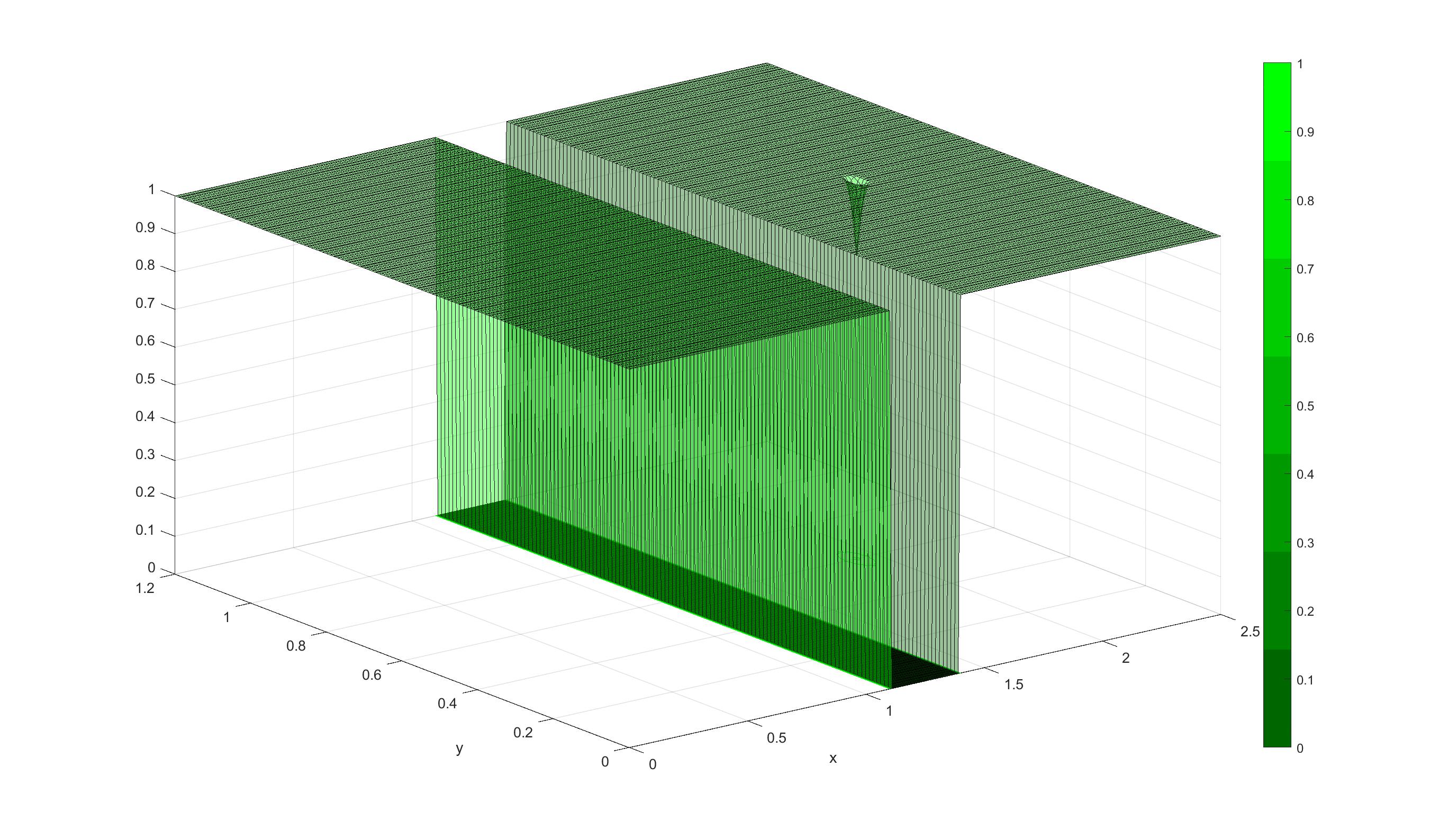}, 
	\includegraphics[width=0.45\textwidth]{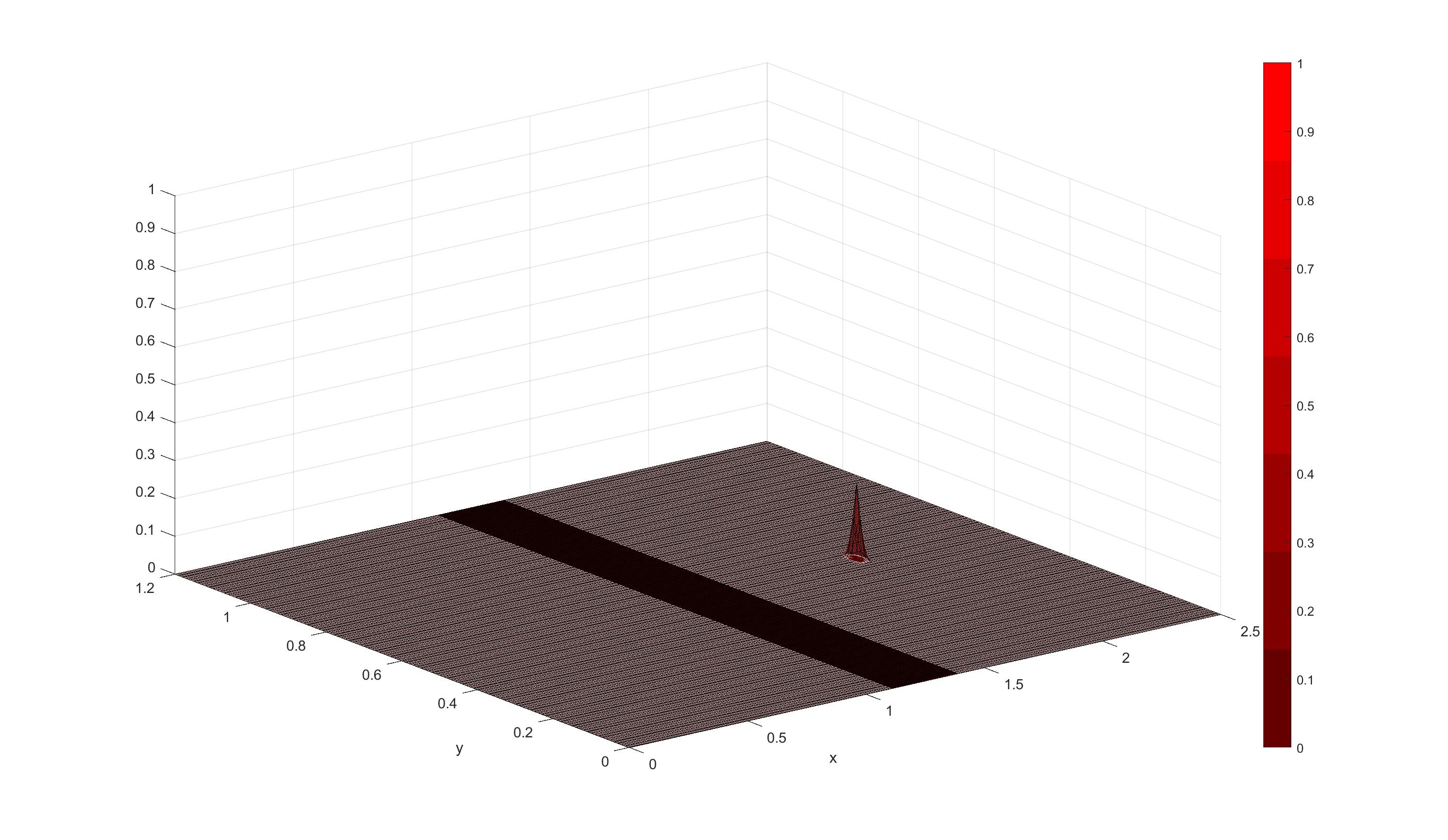}
	\caption{Left: Green trees sub-probability density. Right: Firing trees sub-probability density. Time=0 time u.t. First numerical experiment.}
	\label{fig_i_1}       
\end{center}
\end{figure}

{\begin{figure}
\begin{center}
	\includegraphics[width=0.45\textwidth]{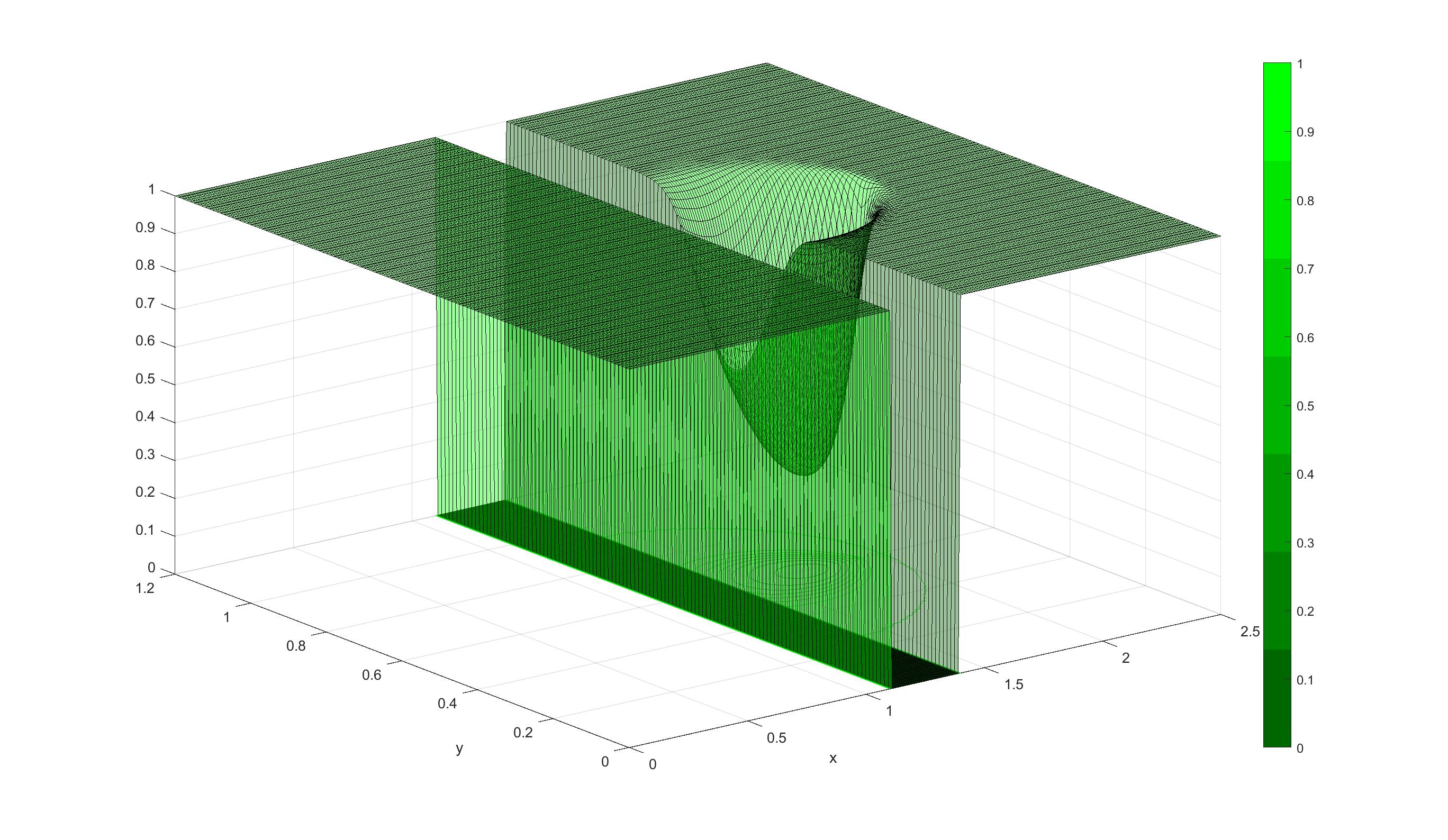}
	\includegraphics[width=0.45\textwidth]{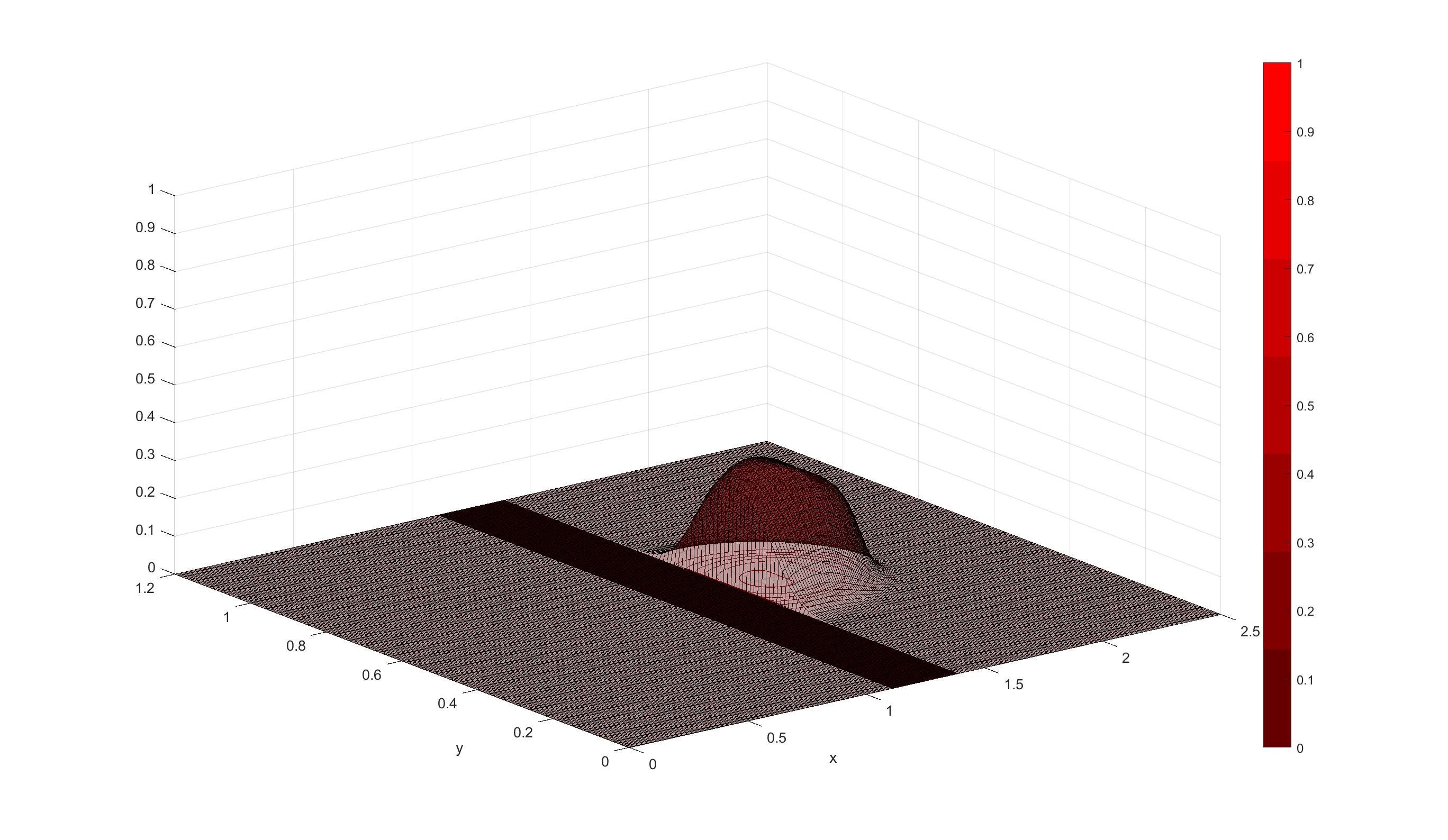}
	\caption{Left: Green trees sub-probability density. Right: Firing trees sub-probability density. Time=0.2 u.t. First numerical experiment.}
	\label{fig_i_2}       
\end{center}
\end{figure}}

{\begin{figure}
\begin{center}
	\includegraphics[width=0.45\textwidth]{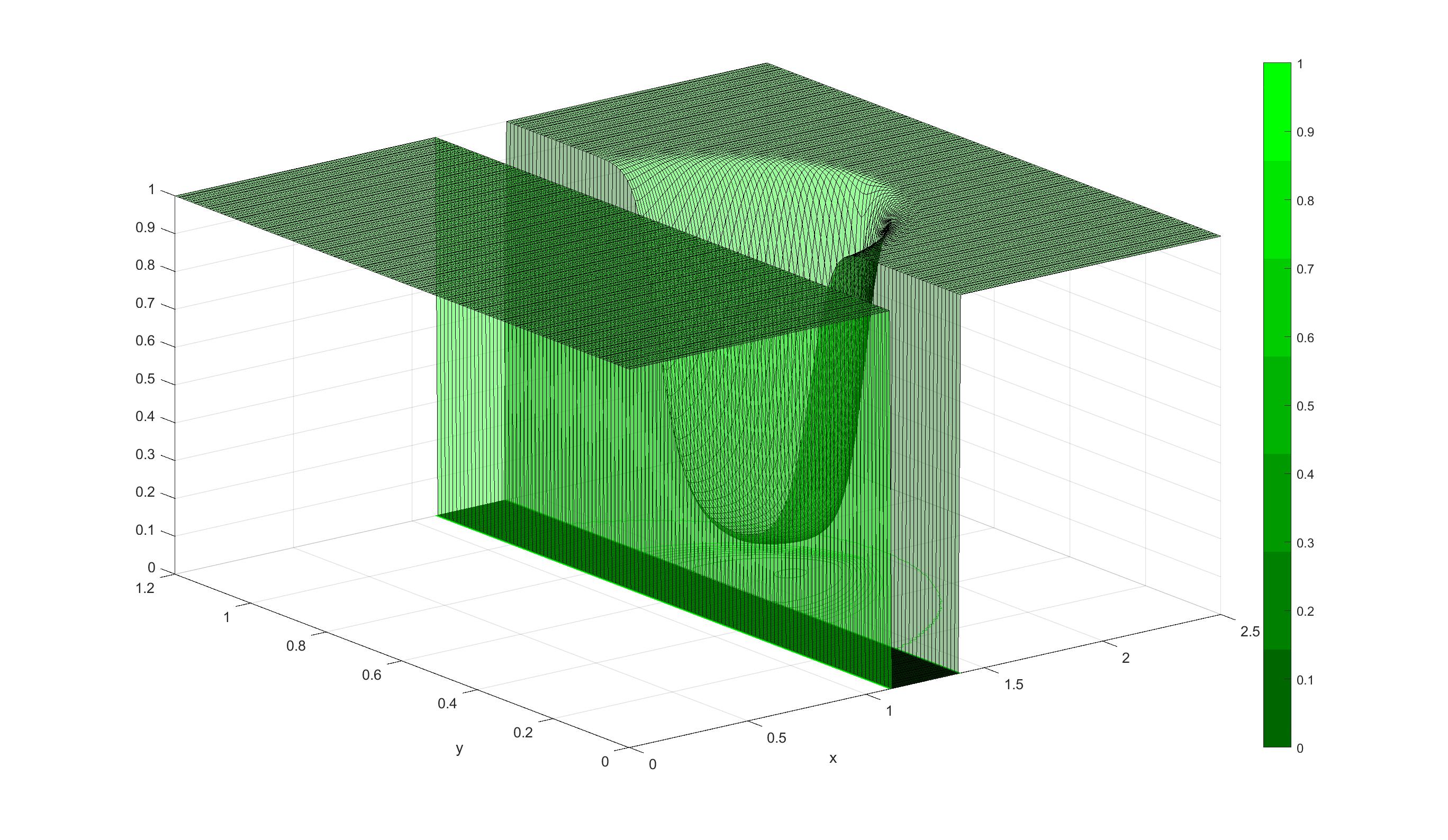}
	\includegraphics[width=0.45\textwidth]{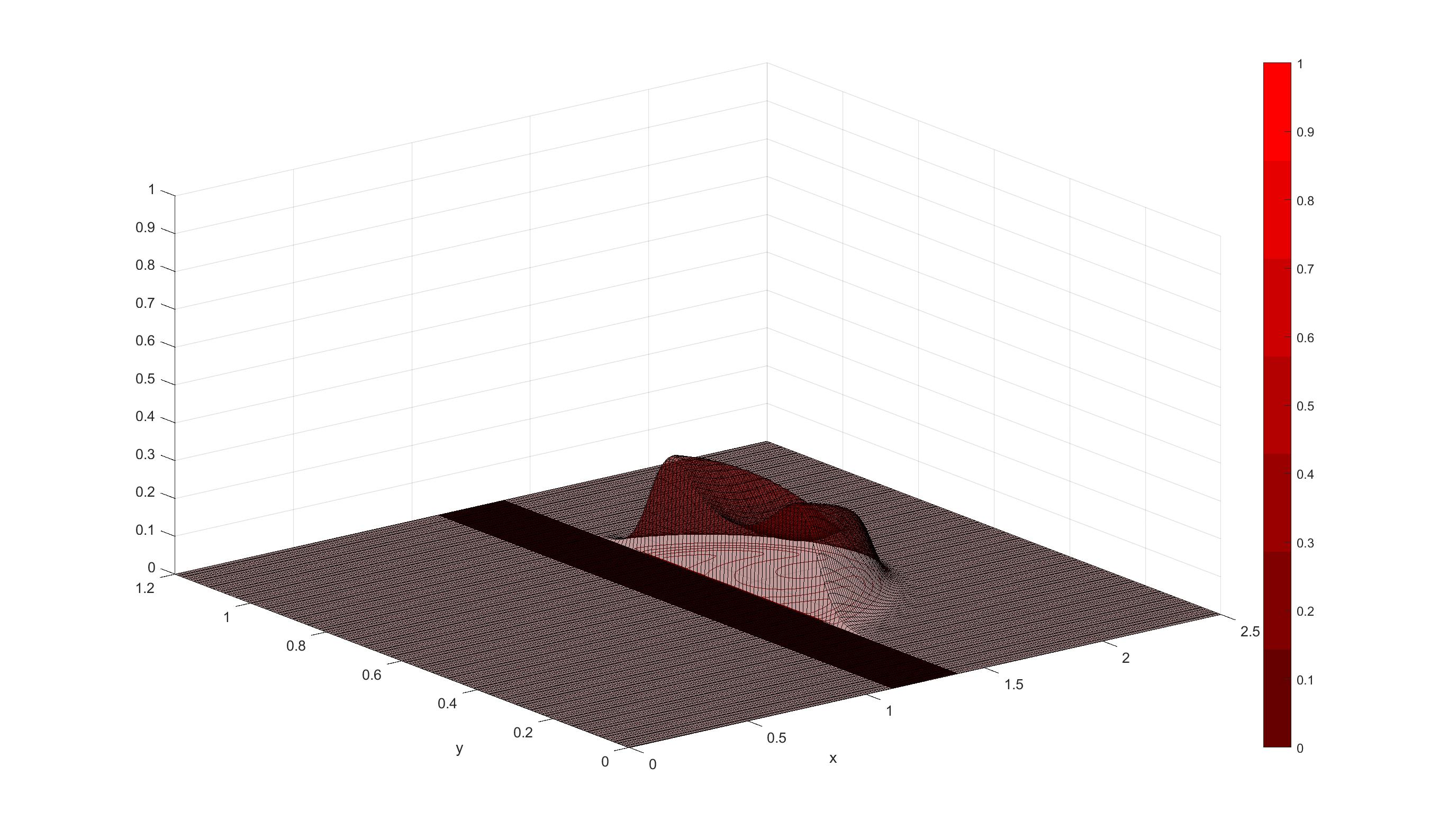}
	\caption{Left: Green trees sub-probability density. Right: Firing trees sub-probability density. Time=0.3 u.t. First numerical experiment.}
	\label{fig_i_3}       
\end{center}
\end{figure}}

{\begin{figure}
\begin{center}
	\includegraphics[width=0.45\textwidth]{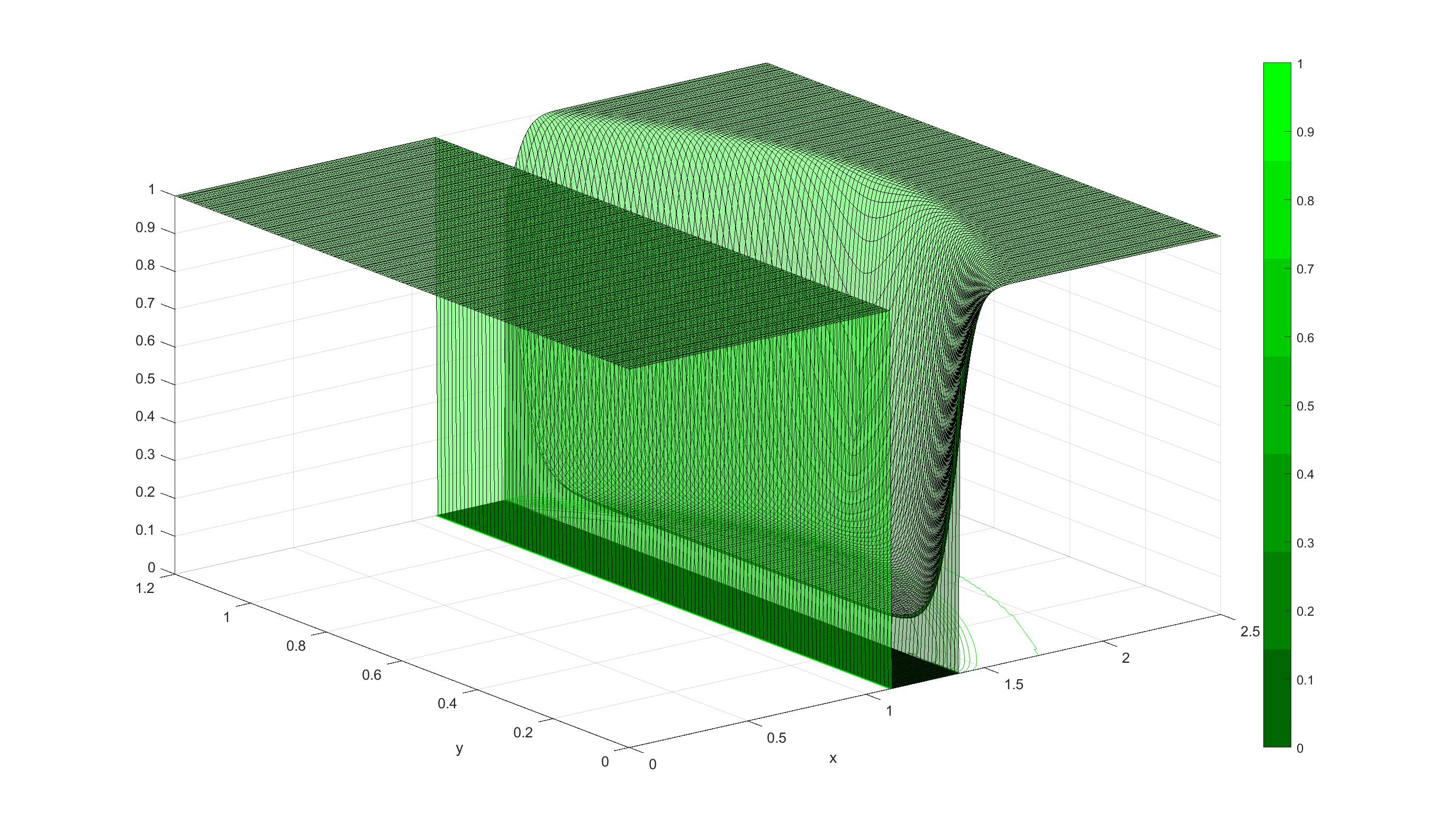}
	\includegraphics[width=0.45\textwidth]{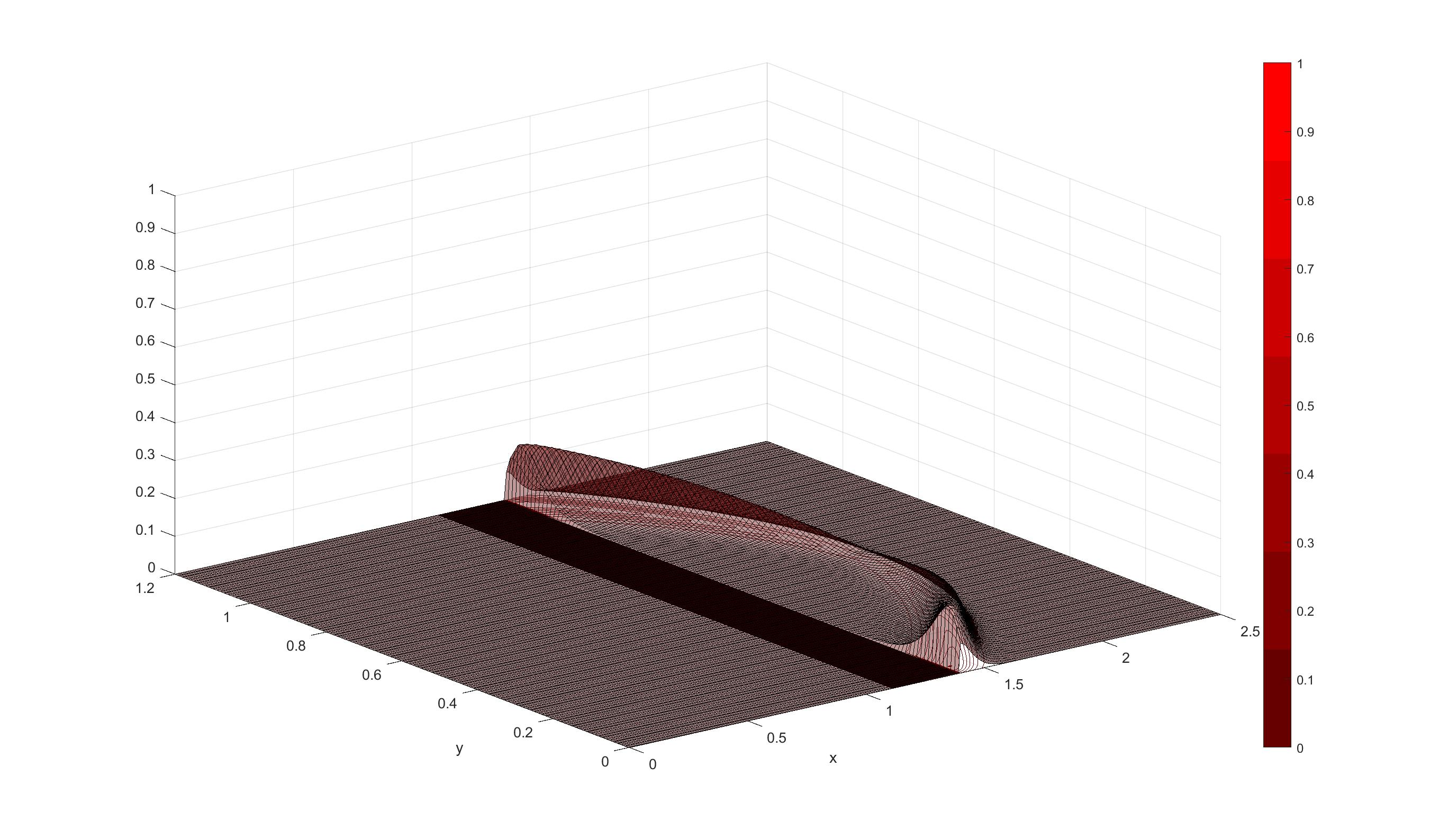}
	\caption{Left: Green trees sub-probability density. Right: Firing trees sub-probability density. Time=1 u.t. First numerical experiment.}
	\label{fig_i_4}       
\end{center}
\end{figure}}

{\begin{figure}
\begin{center}
	\includegraphics[width=0.45\textwidth]{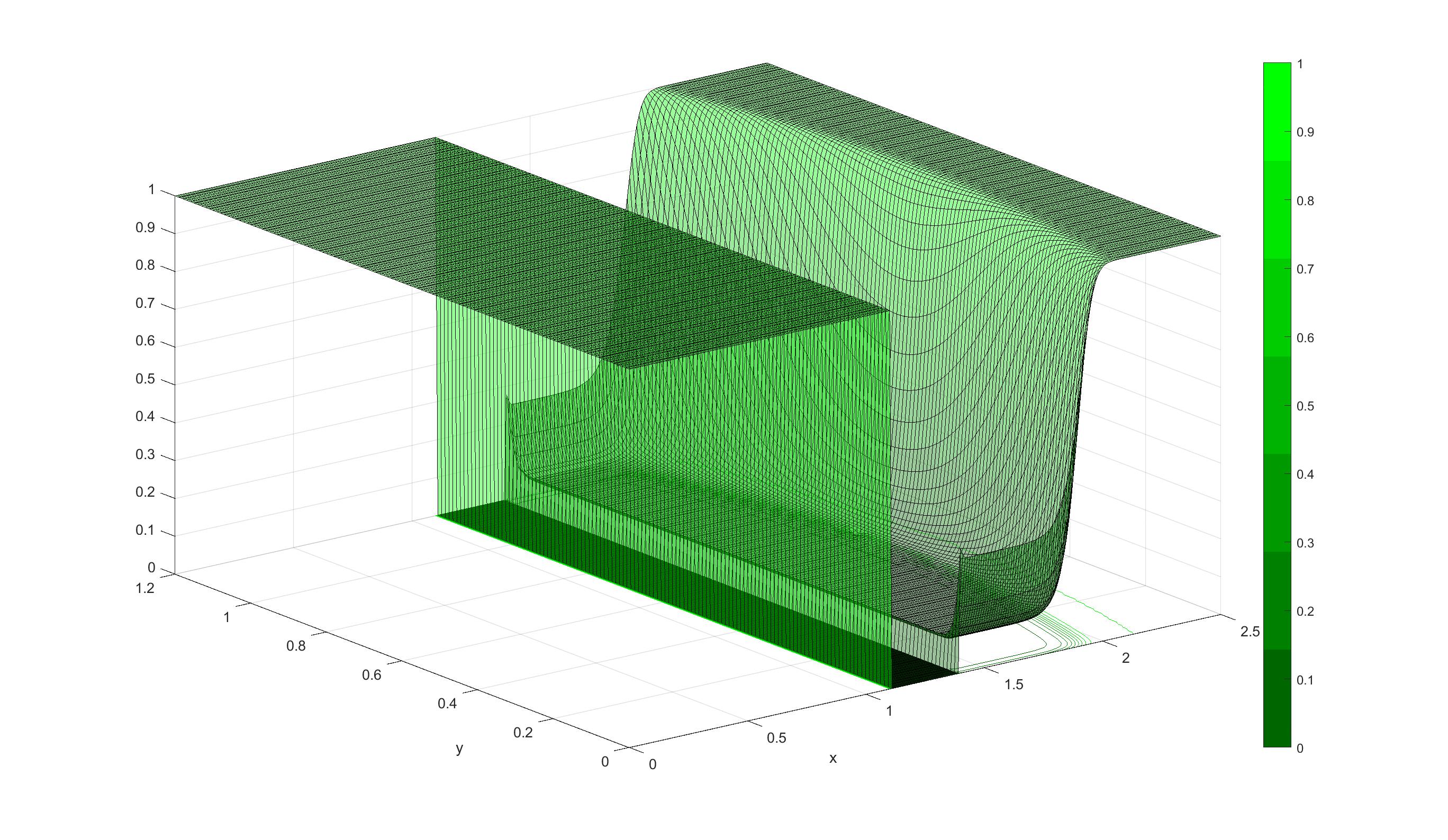}
	\includegraphics[width=0.45\textwidth]{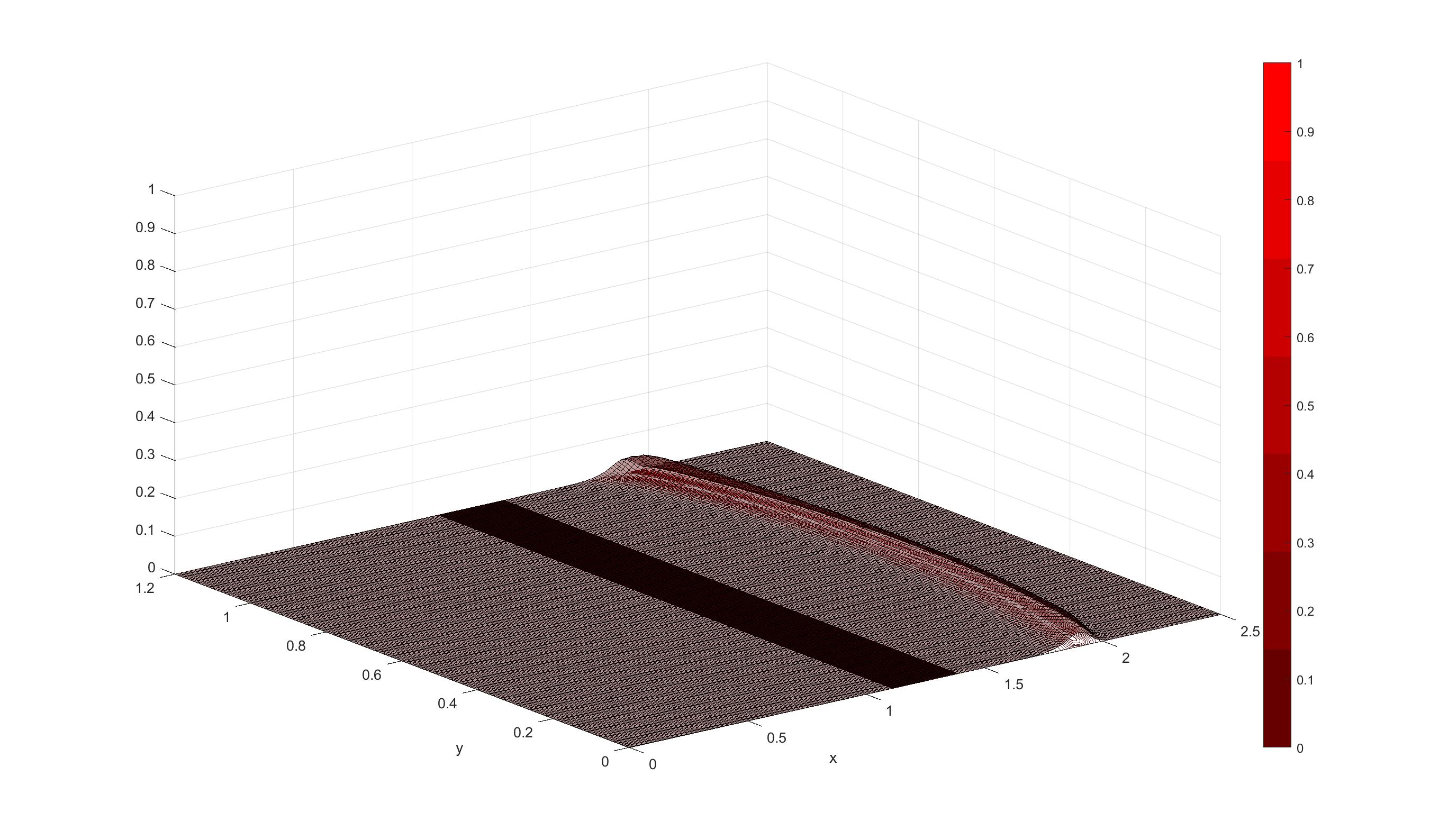}
	\caption{Left: Green trees sub-probability density. Right: Firing trees sub-probability density. Time=3 u.t. First numerical experiment.}
	\label{fig_i_5}       
\end{center}
\end{figure}}

{\begin{figure}
\begin{center}
	\includegraphics[width=0.45\textwidth]{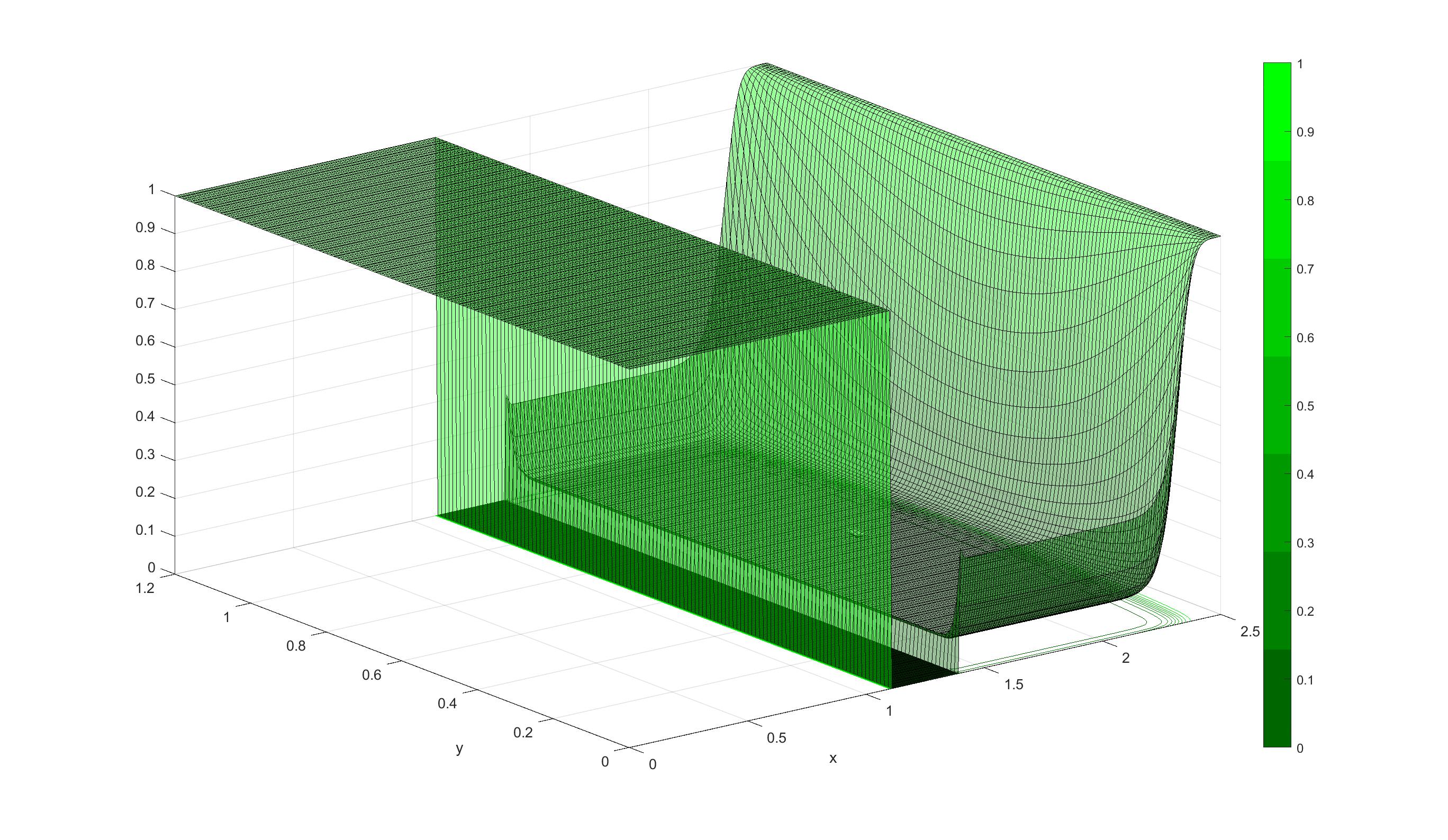}
	\includegraphics[width=0.45\textwidth]{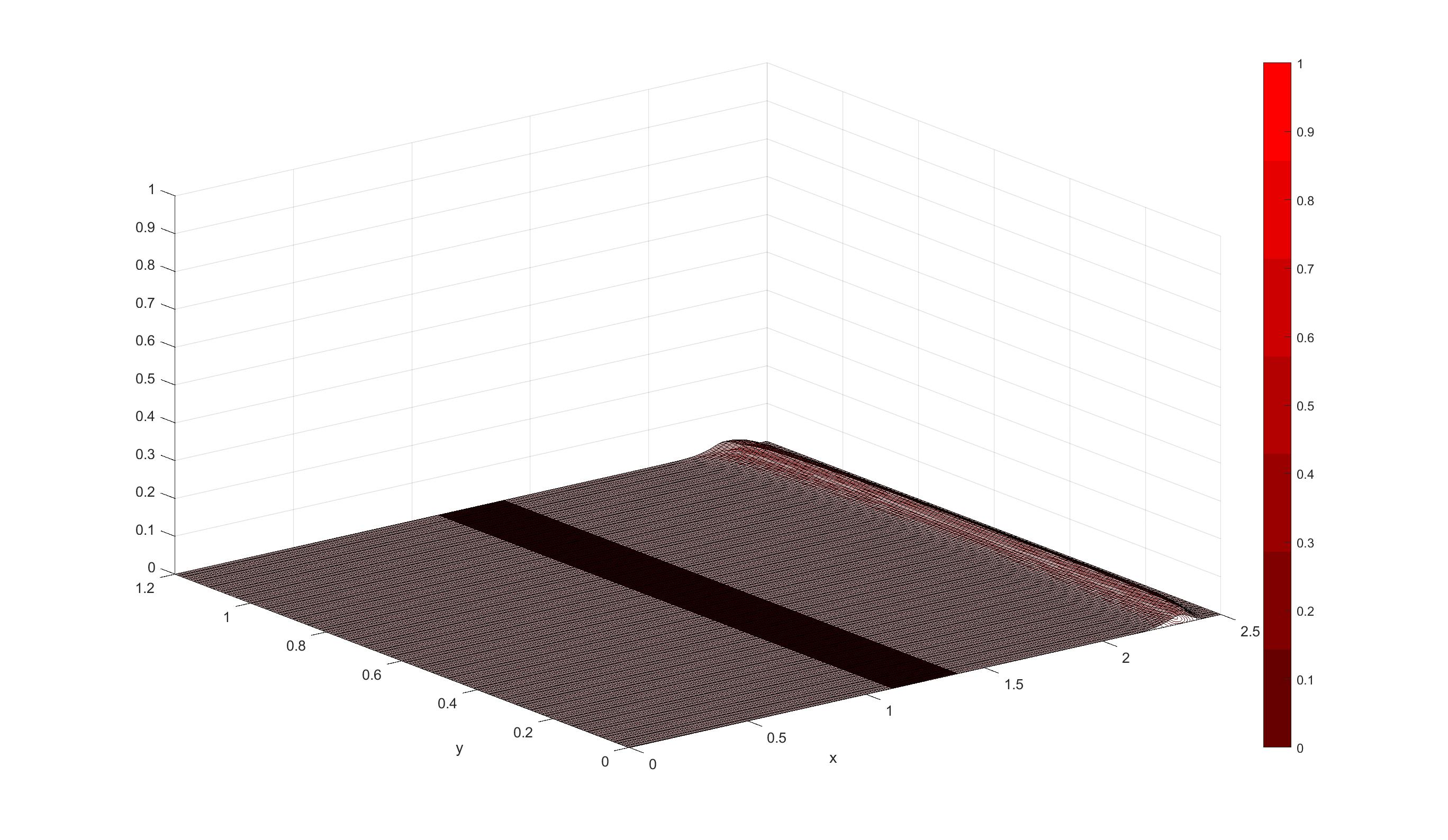}
	\caption{Left: Green trees sub-probability density. Right: Firing trees sub-probability density. Time=7 u.t. First numerical experiment.}
	\label{fig_i_6}       
\end{center}
\end{figure}}

{\begin{figure}
\begin{center}
	\includegraphics[width=0.45\textwidth]{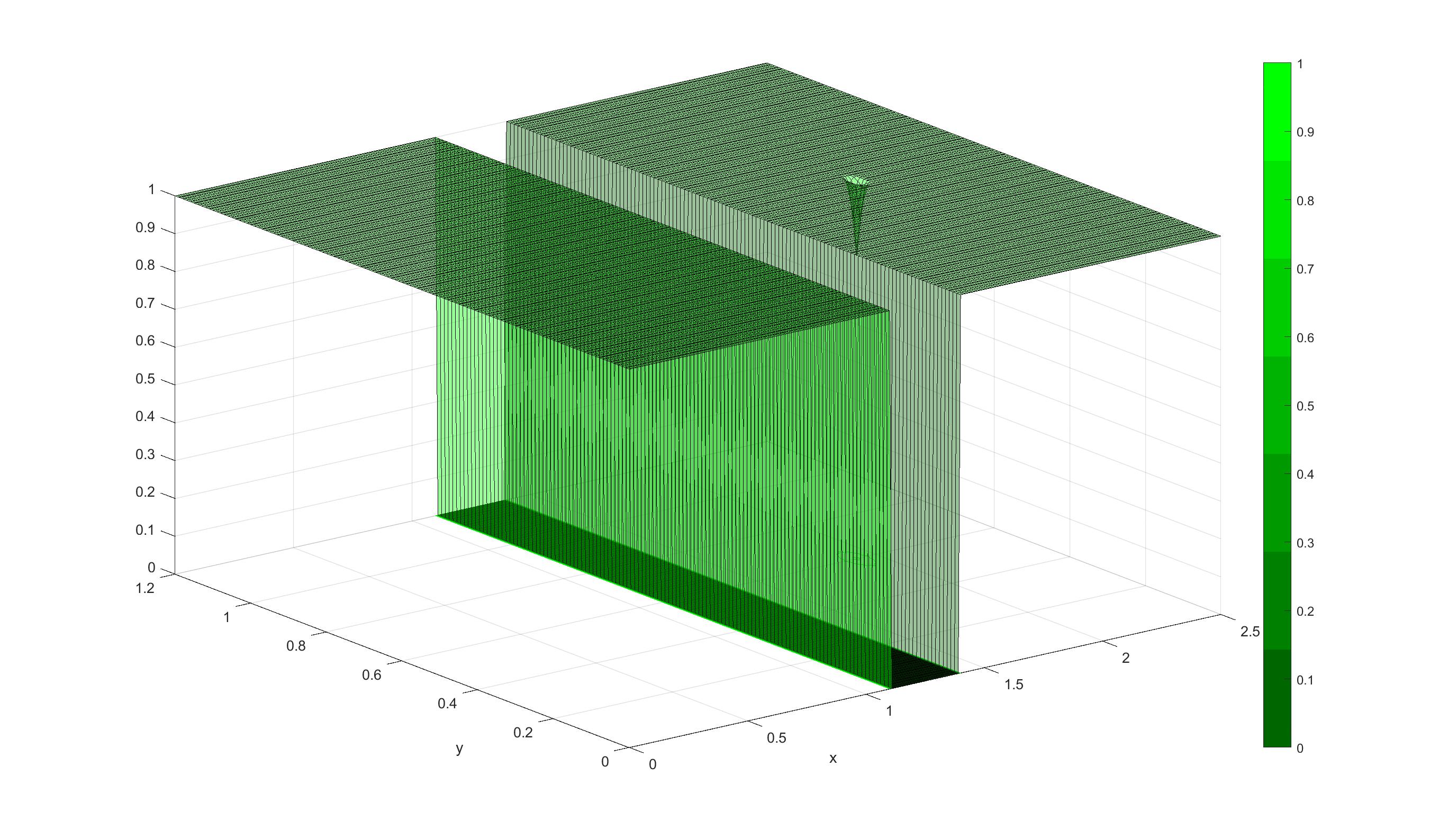}
	\includegraphics[width=0.45\textwidth]{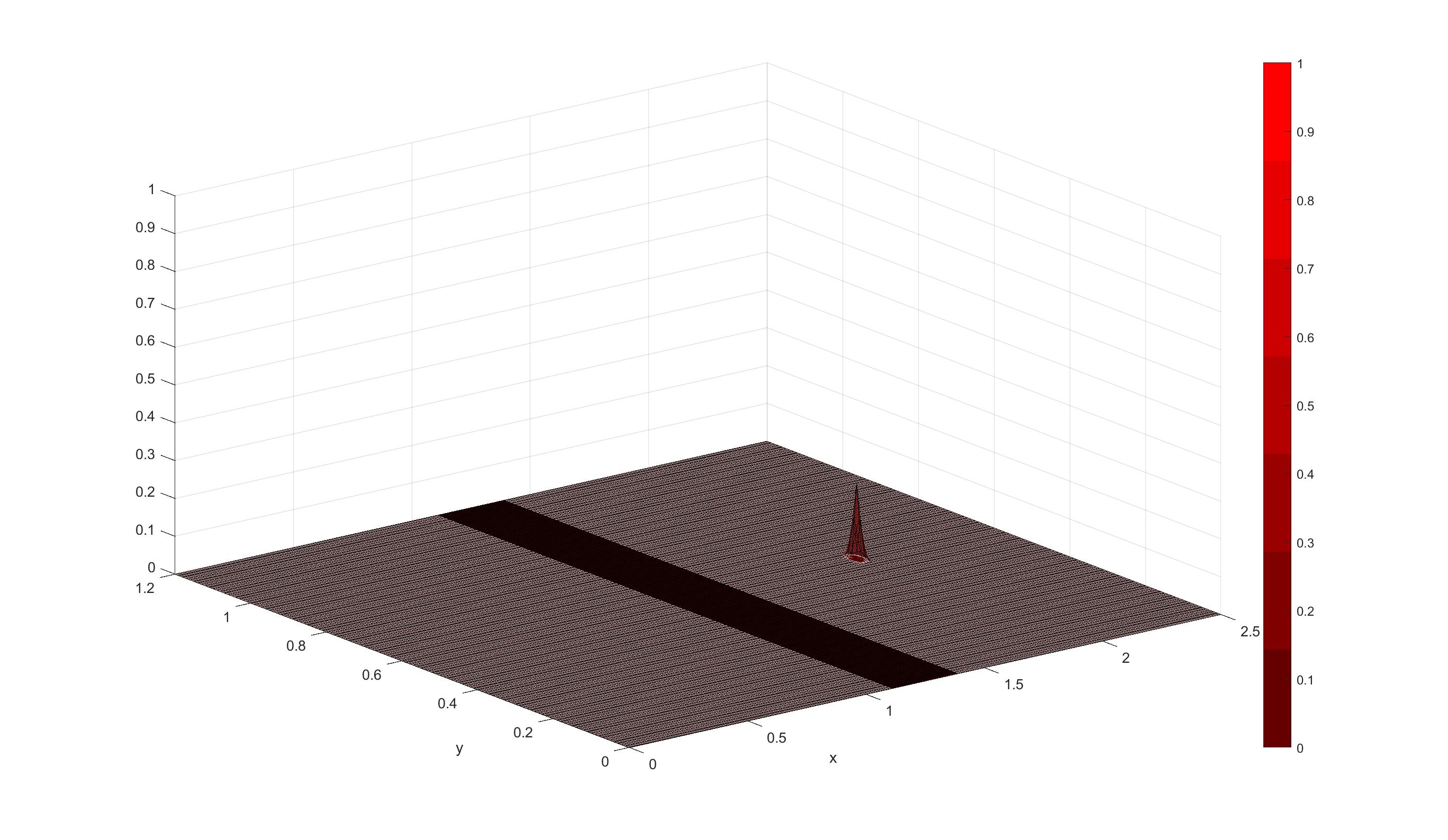}
	\caption{Green trees sub-probability density. Right: Firing trees sub-probability density. Time=0 u.t. Second numerical experiment.}
	\label{fig_i_7}       
\end{center}
\end{figure}}

{\begin{figure}
\begin{center}
	\includegraphics[width=0.45\textwidth]{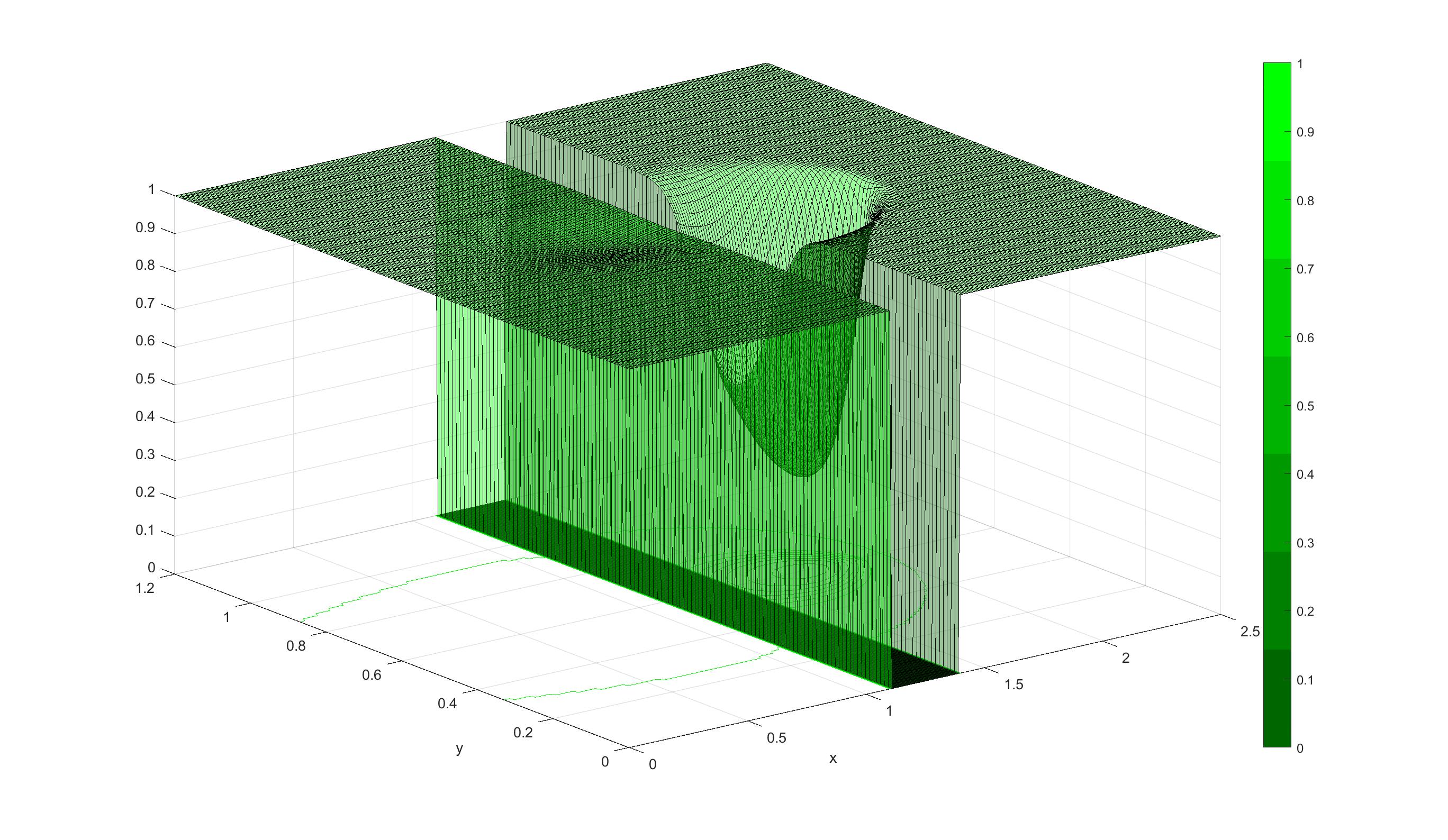}
	\includegraphics[width=0.45\textwidth]{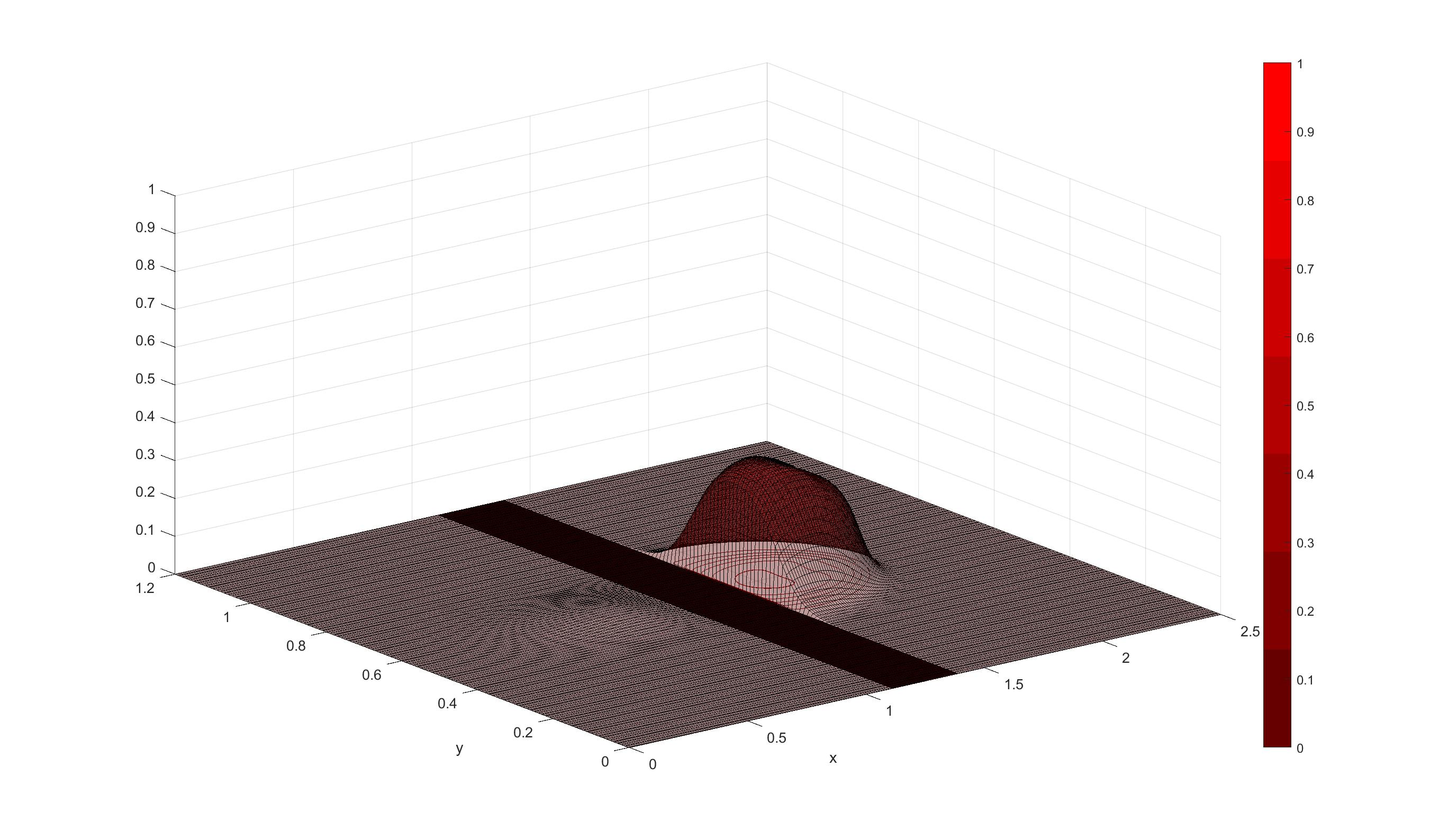}
	\caption{Green trees sub-probability density. Right: Firing trees sub-probability density. Time=0.2 u.t. Second numerical experiment.}
	\label{fig_i_8}       
\end{center}
\end{figure}}

{\begin{figure}
\begin{center}
	\includegraphics[width=0.45\textwidth]{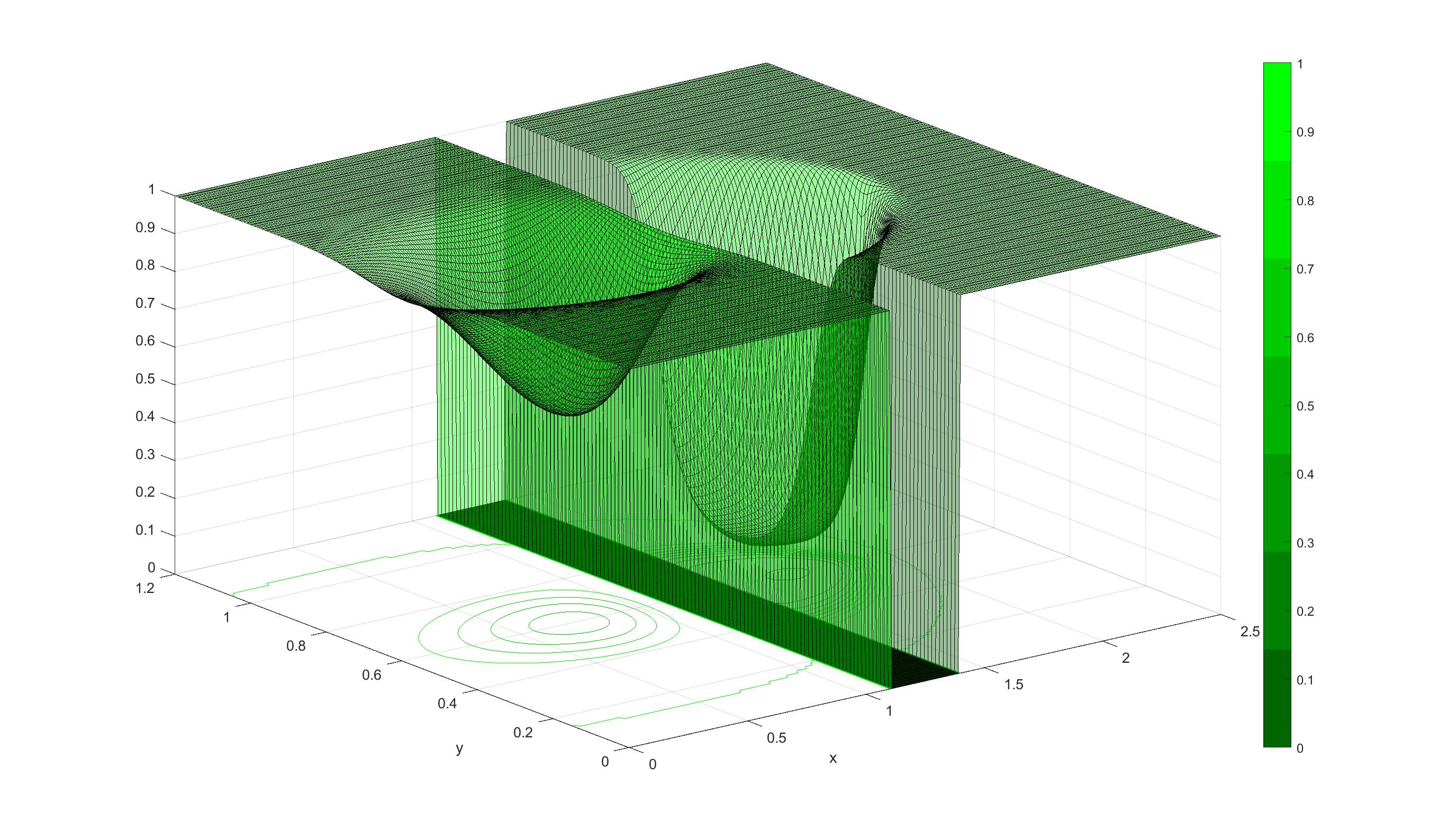}
	\includegraphics[width=0.45\textwidth]{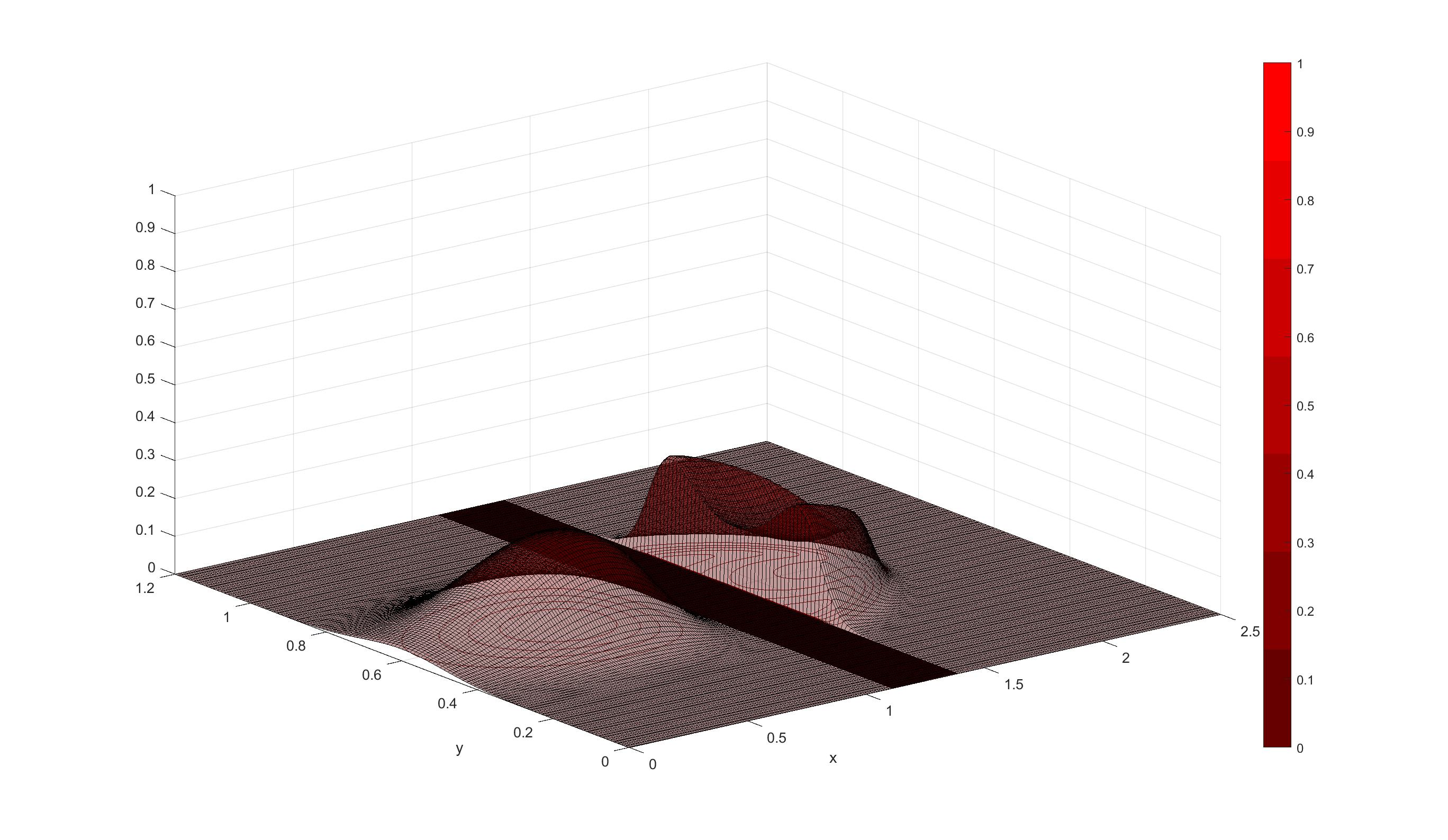}
	\caption{Green trees sub-probability density. Right: Firing trees sub-probability density. Time=0.3 u.t. Second numerical experiment.}
	\label{fig_i_9}       
\end{center}
\end{figure}}

{\begin{figure}
\begin{center}
	\includegraphics[width=0.45\textwidth]{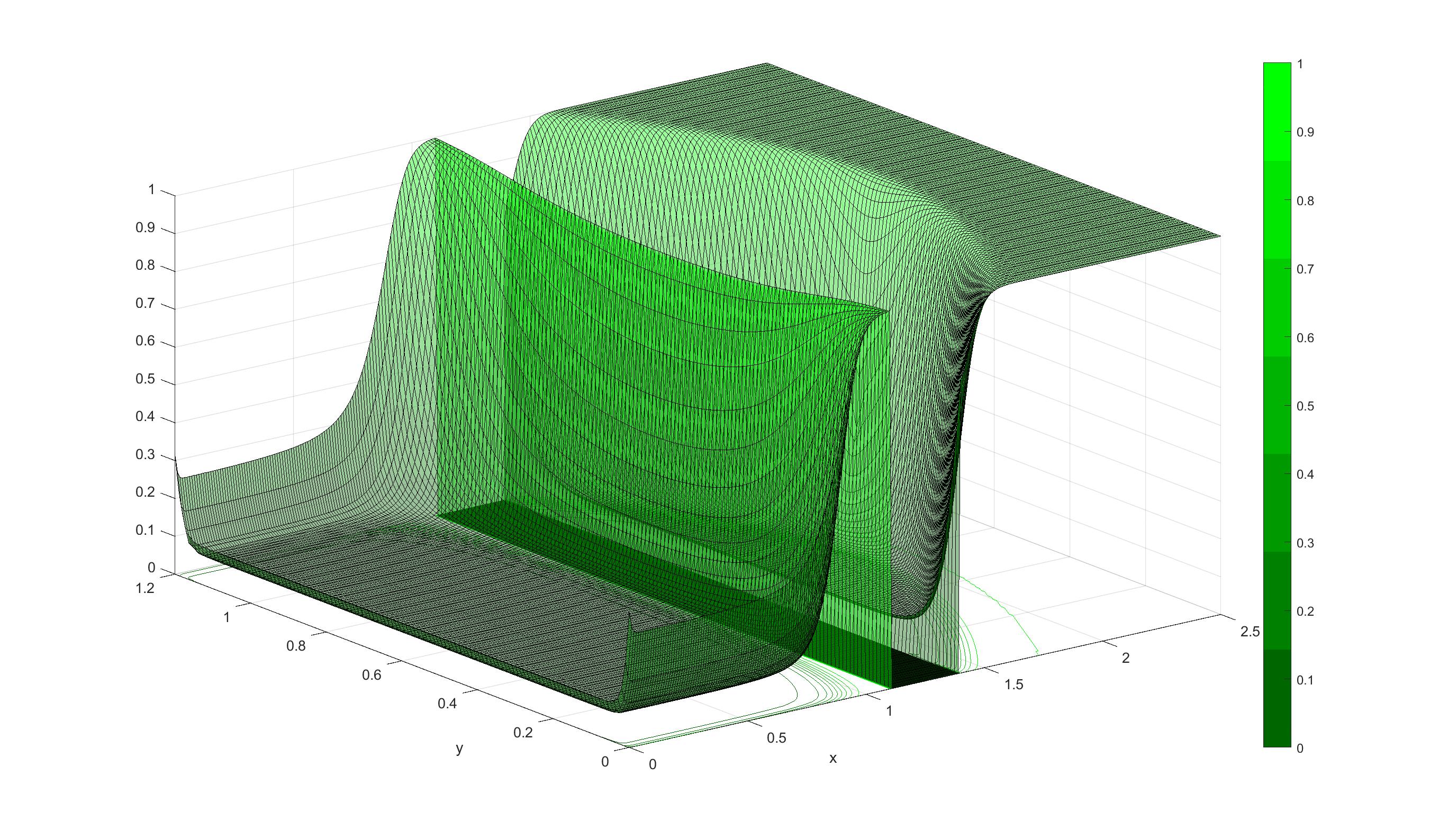}
	\includegraphics[width=0.45\textwidth]{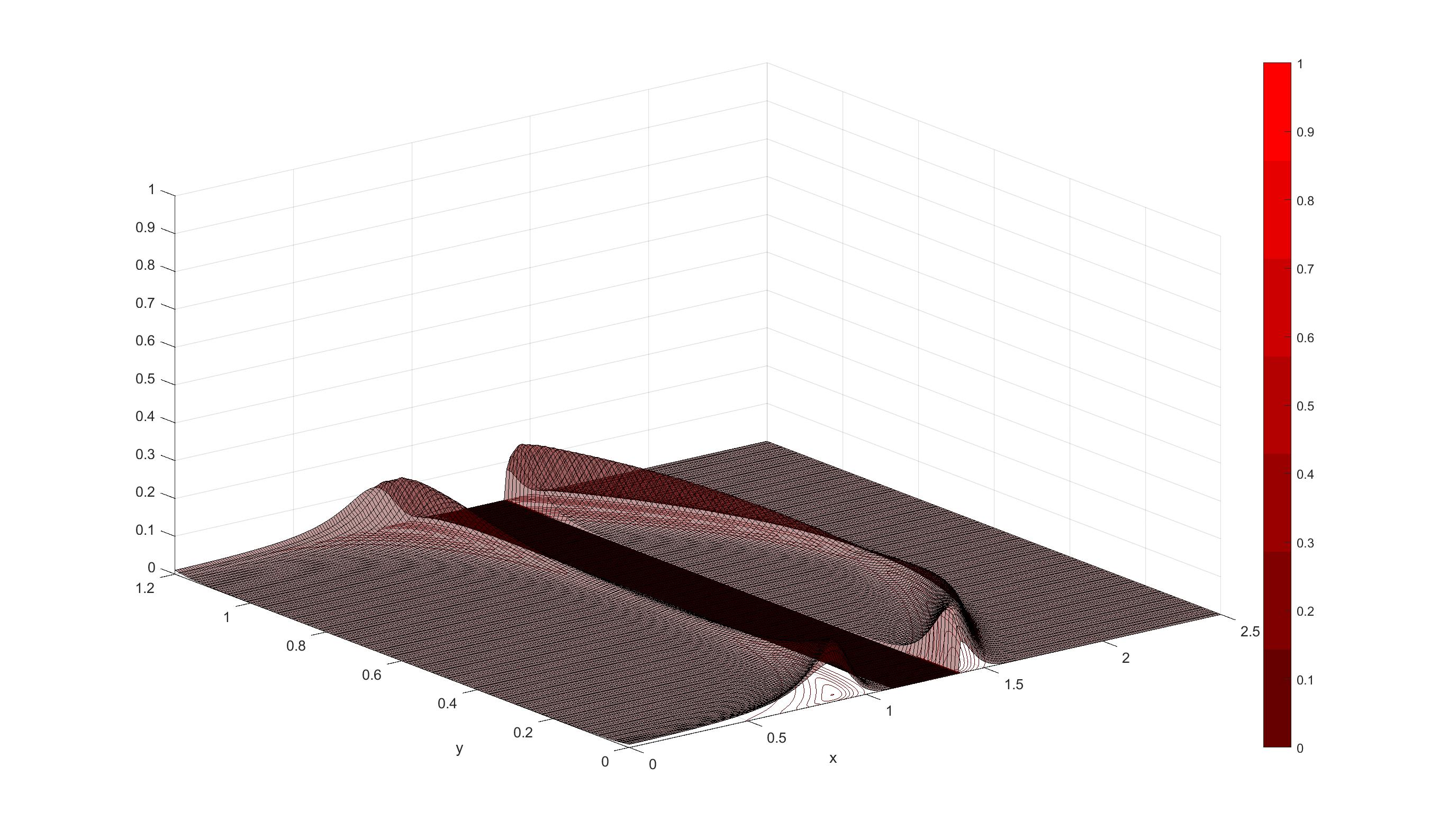}
	\caption{Green trees sub-probability density. Right: Firing trees sub-probability density. Time=1 u.t. Second numerical experiment.}
	\label{fig_i_10}       
\end{center}
\end{figure}}
{\begin{figure}
\begin{center}
	\includegraphics[width=0.45\textwidth]{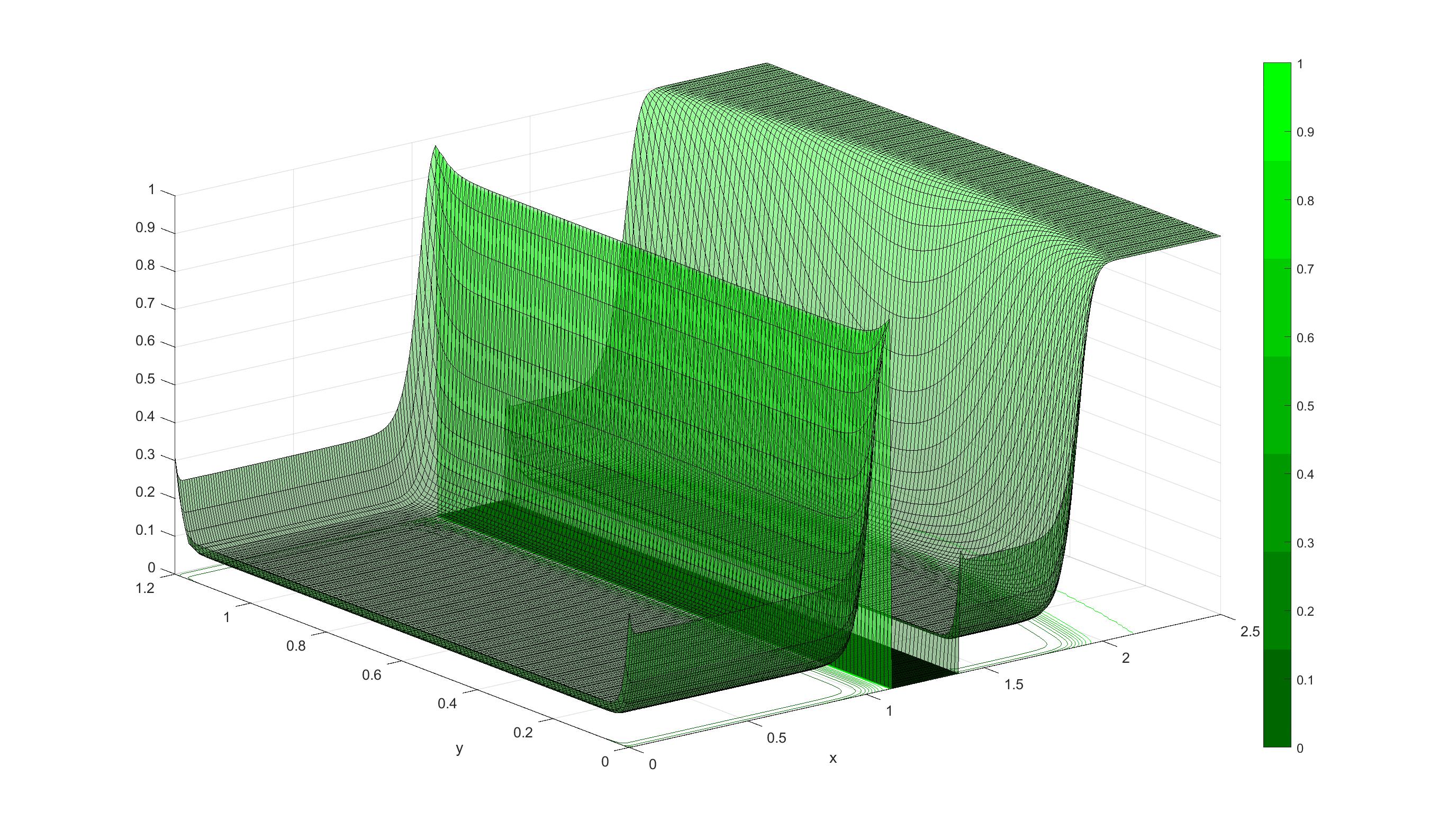}
	\includegraphics[width=0.45\textwidth]{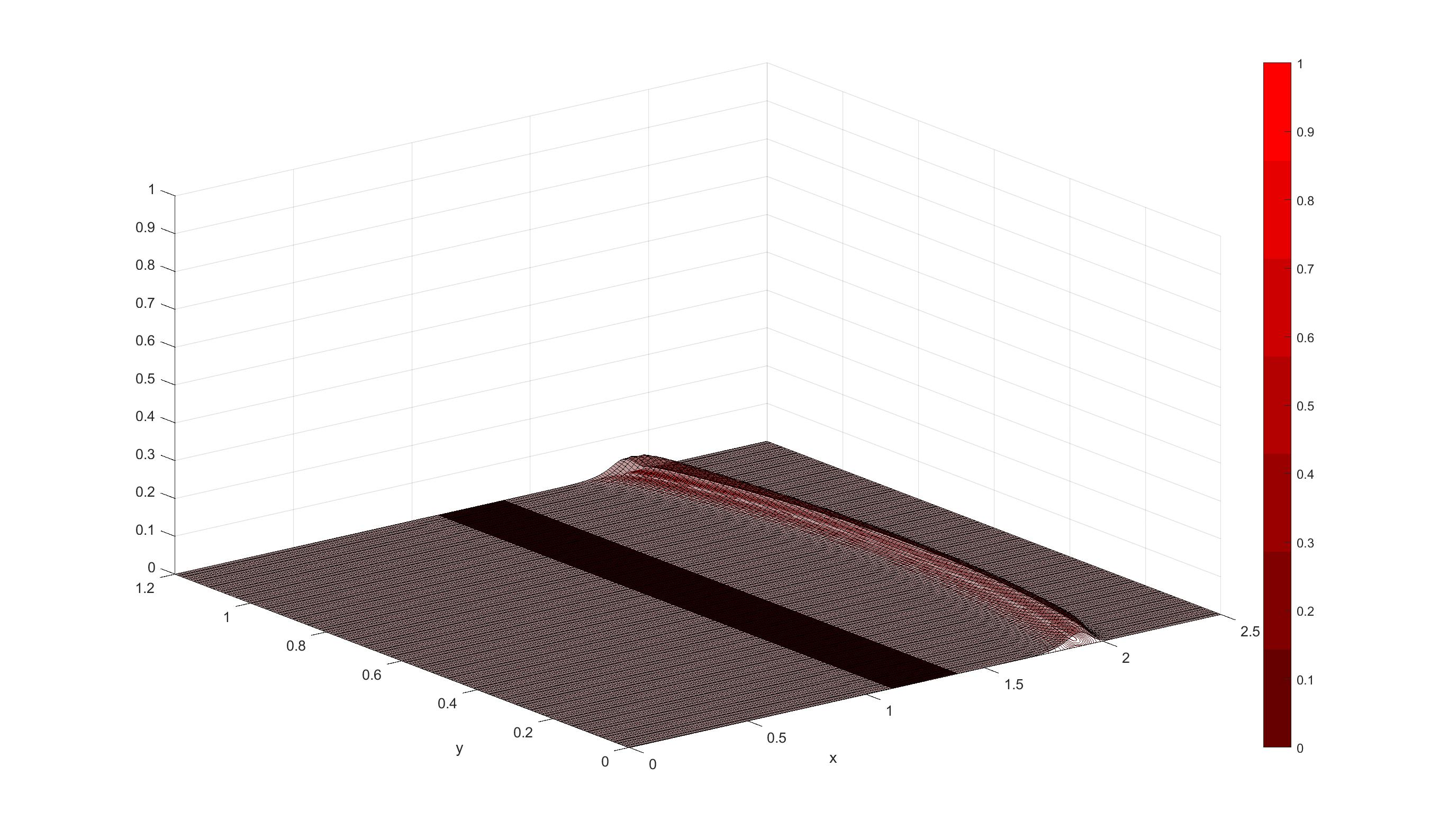}
	\caption{Green trees sub-probability density. Right: Firing trees sub-probability density. Time=3 u.t. Second numerical experiment.}
	\label{fig_i_11}       
\end{center}
\end{figure}}
{\begin{figure}
	\begin{center}
		\includegraphics[width=0.45\textwidth]{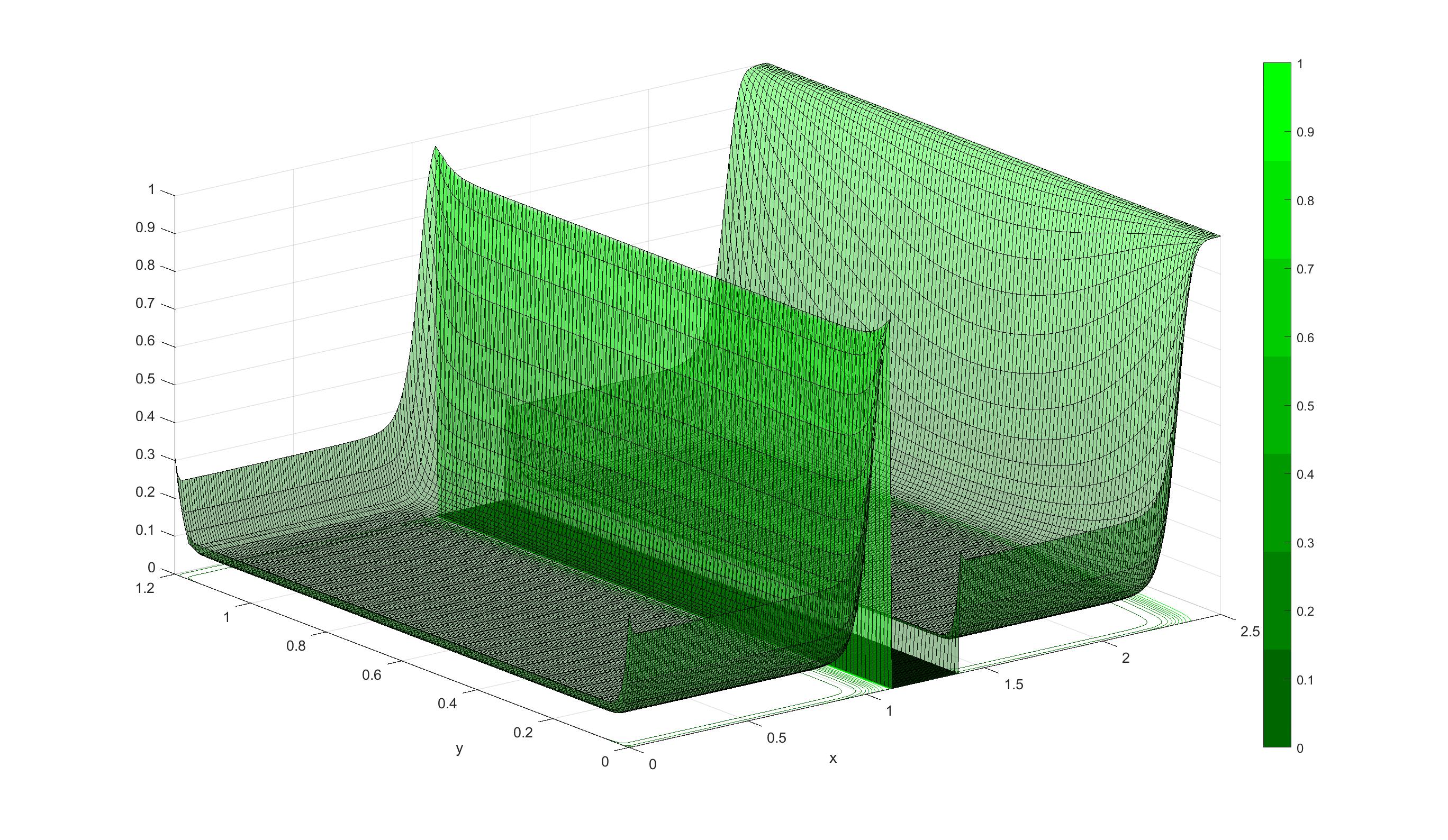}
		\includegraphics[width=0.45\textwidth]{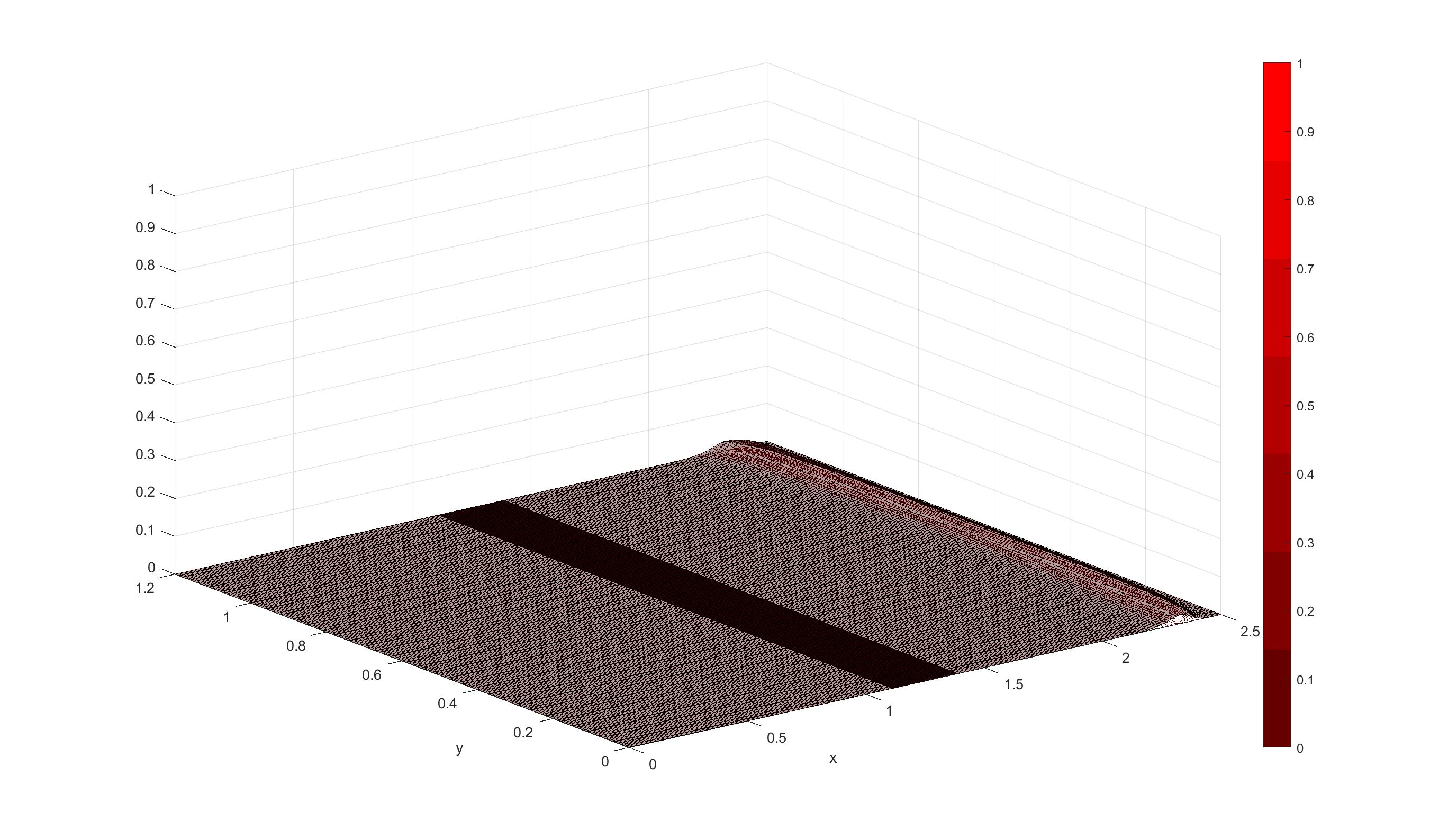}
		\caption{Left: Green trees sub-probability density. Right: Firing trees sub-probability density. Time=7 u.t. Second numerical experiment.}
		\label{fig_i_12}       
	\end{center}
\end{figure}}

\newpage

\section{Conclusions}

We proposed a non-linear stochastic model for forest fire spreading which is continuous in space and time, and can be used in forest fire management. It is well known that forest fires are a contributing factor to climate change and are responsible for large $CO_2$ emissions. Moreover, forest fire spreading is a complex phenomenon characterized by a stochastic behavior whose description requires stochastic models. Until some years ago, continuous stochastic models would have been unhelpful because of the lack of data but nowadays the enormous quantity of geo-referenced data and the availability of powerful techniques for their analysis can provide a very careful picture of forest fires opening the way to more realistic stochastic models that are continuous in space  and time. We provide such a model that is able to exploit geo-referenced data in their full power and can be used for forest fire forecasting. The effects due to land slope, wind and fire spotting are included. Moreover, the model can take into account the effects of firefighters interventions. Finally,  at variance with stochastic models based on cellular automata where the spreading happens only in few directions, it is continuous in space. In our knowledge it is the first model of this kind. Besides, it is a kind of mother model from which many other models (both stochastic and deterministic) can be derived. For example, stochastic models based on cellular automata can be derived by time and/or space discretization. 

The state of the forest fire is described by the sub-probability densities of green and firing trees (both can be estimated thanks to data coming from satellites and earth detectors) whose evolution is determined by a density kernel $W$ and can be calculated by solving a system of integro-differential equations. The evolution of the sub-probability densities predicted by the model can be used to locate those regions in space where the fire probability is going to increase, providing crucial information for firefighters. 
 
In order to implement the model numerically we used a phenomenological approach inspired by some theoretical results \cite{Sardoy,Perry} that suggested a particular structure for $W$. By using fourth order Runge-Kutta scheme we realized two numerical experiments and calculated the evolution of the probability densities in the case of 1) a forest fire propagating towards a river in the presence of wind and 2) a forest fire propagating towards a river in the presence of both wind and fire spotting. The simulation shows that in the first case the fire propagation stops in correspondence to the river while in the second case the probability that the fire propagates beyond the river is different from zero. In both cases the probability of downwind propagation is different from zero but lower than the probability of propagation in the wind direction.

\section*{Acknowledgements}

\noindent
The present work was performed under the auspices of the GNFM (Gruppo Nazionale di Fisica Matematica).

\noindent
The present research has been funded by the European Union and by the Italian Ministry of Research (NextGenerationEU), project title: TECH4YOU, SPOKE 1 GOAL 1.4 PP3, project number: $ECS-00000009$. 																															 
\newpage

\section*{Appendix}

In this Appendix, we study the existence and uniqueness of solutions for the system (\ref{system}) and for the system defined by equations (\ref{evolutionF}) and (\ref{evolutionG1}). In particular, by using the Banach fixed point theorem, we prove the existence and uniqueness of solutions of the two systems. For the reader convenience we copy the two systems below,

\begin{equation}\label{system1}
	\begin{cases}
		&\psi^F(t,\mathbf{x})=\psi^F(0,\mathbf{x})-\psi^B(t,\mathbf{x})+\int_{\Sigma}\int_0^{t}\,\psi^G(\tau,\mathbf{x})W(\tau,\mathbf{x},\mathbf{y})\vert\psi^F(\tau,\mathbf{y})\vert\,d\tau\,d^2\mathbf{y}\\
		&\psi^G(t,\mathbf{x})=\psi^G(0,\mathbf{x})-\int_{\Sigma}\int_0^{t}\,\psi^G(\tau,\mathbf{x})W(\tau,\mathbf{x},\mathbf{y})\vert\psi^F(\tau,\mathbf{y})\vert\,d\tau\,d^2\mathbf{y}
	\end{cases}
\end{equation}
and

\begin{equation}
	\label{evolutionFG}
	\begin{cases}
	&\psi^F(t,\mathbf{x})=\psi^F(0,\mathbf{x})\pi(t)+\int_0^{t}\,\int_{\Sigma}\psi^G(\tau,\mathbf{x})W(\tau,\mathbf{x},\mathbf{y})
	\psi^F(\tau,\mathbf{y})\pi(t-\tau)\,d\tau\,d^2\mathbf{y}\\
	&\psi^G(t,\mathbf{x})=\psi^G(0,\mathbf{x})-\int_0^t\int_\Sigma\psi^G(\tau,\mathbf{x})W(\tau,\mathbf{x},\mathbf{y})\psi^F(\tau,\mathbf{y})\,d\tau\,d^2\mathbf{y}.
	\end{cases}
\end{equation}
for the memoryless and the memory cases respectively. 
\mbox{ } \\[1cm]

\subsection*{Existence and uniqueness of solutions I}

In order to ensure that system (\ref{system1}) admits a unique solution, we consider the operator $\hat{F}:L^\infty(\mathbb{R}\times\Sigma,dt\,d^2\mathbf{x})\oplus L^\infty(\mathbb{R}\times\Sigma,dt\,d^2\mathbf{x})\to L^\infty(\mathbb{R}\times\Sigma,dt\,d^2\mathbf{x})\oplus L^\infty(\mathbb{R}\times\Sigma,dt\,d^2\mathbf{x})$ defined by

\begin{equation}\label{S}
	\begin{pmatrix}
		\widetilde{\psi}^F(t,\mathbf{x})\\
		\widetilde{\psi}^G(t,\mathbf{x})\\
	\end{pmatrix}:=\hat{F}\begin{pmatrix}
		\psi^F\\
		\psi^G\\
	\end{pmatrix}(t,\mathbf{x})=
	\begin{pmatrix}
		\psi^F(0,\mathbf{x})-\psi^B(t,\mathbf{x})+\int_{\Sigma}\int_0^{t}\,\psi^G(\tau,\mathbf{x})W(\tau,\mathbf{x},\mathbf{y})\vert\psi^F(\tau,\mathbf y)\vert\,d\tau\,d^2\mathbf{x}\\
		\psi^G(0,\mathbf{x})-\int_{\Sigma}\int_0^{t}\,\psi^G(\tau,\mathbf{x})W(\tau,\mathbf{x},\mathbf{y})\vert\psi^F(\tau,\mathbf y)\vert\,d\tau\,d^2\mathbf{y}\\
	\end{pmatrix}
\end{equation}

\noindent 
and make the following assumptions. For all $t\in[0,\infty)$: 
\begin{itemize}
	\item[1)] we assume $\psi^G(t,\mathbf{x}),\psi^F(t,\mathbf{x})\in L^\infty(\mathbb{R}\times\Sigma,dt\,d^2\mathbf{x})$ to be continuous for all $\mathbf{x}\in\Sigma$,
	\item[2)] $0\leq\psi^G(t,\mathbf{x})\leq\psi^G(0,\mathbf{x})\leq M$, $\vert\psi^F(t,\mathbf{x})\vert\leq2M$,  
	$\psi^F(0,\mathbf{x})\geq 0$ for all $\mathbf{x}\in\Sigma$, $\psi^G(0,\mathbf{x})+\psi^F(0,\mathbf{x})\leq M$, where $M$ is a positive constant, 
	\item[3)] concerning the kernel $W$, we assume $0\leq W(t,\mathbf{x},\mathbf{y})\leq\bar{W}\in\mathbb{R}_+$. Moreover, we assume that there is a $T>0$ such that $\int_{kT}^{(k+1)T}\alpha(\tau)\,d\tau\leq\frac{1}{2}$, and $2M|\Sigma|\bar{W}T\leq\frac{1}{4}$, for every $k\in\mathbb{N}$, where $|\Sigma|$ denotes the area of the region $\Sigma$.
\end{itemize} 


\noindent

\noindent
The space $\mathcal{L}^\infty(\mathbb R):=L^\infty(\mathbb{R}\times\Sigma,dt\,d^2\mathbf{x})\oplus L^\infty(\mathbb{R}\times\Sigma,dt\,d^2\mathbf{x})$ with the norm 

$$\|(\psi^F,\psi^G)\|_\infty=\max\{\|\psi^F\|_\infty,\|\psi^G\|_\infty\}$$

\noindent
is a Banach space.

The proof is divided into steps. Let us consider the space $S_{nT}$ which is the subset of $L^\infty([(n-1)T,nT]\times\Sigma,dt\,d^2\mathbf{x})\oplus L^\infty([(n-1)T,nT]\times\Sigma,dt\,d^2\mathbf{x})$, with $n=1,2,\dots$, such that the following assumptions hold 
\begin{itemize}
	\item[1)]  $\psi^G(t,\mathbf{x}),\psi^F(t,\mathbf{x})\in L^\infty([(n-1)T,nT]\times\Sigma,dt\,d^2\mathbf{x})$  continuous for all $\mathbf{x}\in\Sigma$,
	\item[2)] $0\leq\psi^G(t,\mathbf{x})\leq\psi^G((n-1)T,\mathbf{x})\leq M$, $\vert\psi^F(t,\mathbf{x})\vert\leq2M$,  
	$\psi^F((n-1)T,\mathbf{x})\geq 0$ for all $\mathbf{x}\in\Sigma$, $\psi^G((n-1)T,\mathbf{x})+\psi^F((n-1)T,\mathbf{x})\leq M$, where $M$ is a positive constant.
\end{itemize} 

Moreover, we introduce $g(\mathbf{x}):=\psi^G(0,\mathbf{x})$, $f(\mathbf{x}):=\psi^F(0,\mathbf{x})$ (that correspond to the choice of the initial condition for the system). 

Let $\hat{F}_T:L^\infty([0,T]\times\Sigma,dt\,d^2\mathbf{x})\oplus L^\infty([0,T]\times\Sigma,dt\,d^2\mathbf{x})\to L^\infty([0,T]\times\Sigma,dt\,d^2\mathbf{x})\oplus L^\infty([0,T]\times\Sigma,dt\,d^2\mathbf{x})$ be the operator defined in the time interval $[0,T]$ by

\begin{align}\label{S1}
	\begin{pmatrix}
		\widetilde{\psi}^F(t,\mathbf{x})\\
		\widetilde{\psi}^G(t,\mathbf{x})\\
	\end{pmatrix}:&=\hat{F}_T\begin{pmatrix}
		\psi^F\\
		\psi^G\\
	\end{pmatrix}(t,\mathbf{x})\notag\\
	&=\begin{pmatrix}
		\psi^F(0,\mathbf{x})-\psi^B(t,\mathbf{x})+\int_{\Sigma}\int_0^{t}\,\psi^G(\tau,\mathbf{x})W(\tau,\mathbf{x},\mathbf{y})\vert\psi^F(\tau,\mathbf{y})\vert\,d\tau\,d^2\mathbf{x}\\
		\psi^G(0,\mathbf{x})-\int_{\Sigma}\int_0^{t}\,\psi^G(\tau,\mathbf{x})W(\tau,\mathbf{x},\mathbf{y})\vert\psi^F(\tau,\mathbf y)\vert\,d\tau\,d^2\mathbf{y}\\
	\end{pmatrix}.
\end{align}

In order for Banach fixed point theorem to be applied, we need to show that $\hat{F}_TS_{T}\subset S_{T}$ and that $\hat{F}_T:S_{T}
\to S_{T}$ is a contraction.
First, we show that $\hat{F}_T S_{T}\subset S_{T}$. By conditions 1), 2) and 3), we have

\begin{align*}
	&\int_{\Sigma}\int_0^{t}\,\psi^G(\tau,\mathbf{x})W(\tau,\mathbf{x},\mathbf{y})\vert\psi^F(\tau,\mathbf{y})\vert\,d\tau\,d^2\mathbf{y}\leq\psi^G(0,\mathbf{x})\int_{\Sigma}\int_{0}^{t}\, W(\tau,\mathbf{x},\mathbf{y})\vert\psi^F(\tau,\mathbf{y})\vert\,d\tau\,d^2\mathbf{y}\\
	&\leq\psi^G(0,\mathbf{x})2M\Sigma\bar{W} T\leq\psi^G(0,\mathbf{x}).
\end{align*}

\noindent
Therefore,

$$0\leq\widetilde{\psi}^G(t,\mathbf{x})\leq\widetilde{\psi}^G(0,\mathbf{x})=\psi^G(0,\mathbf{x})\leq M.$$

\noindent
We proceed analogously for $\widetilde{\psi}^F$,

\begin{align*} 
	\vert\widetilde{\psi}^F(t,\mathbf{x})\vert&=\Big\vert\psi^F(0,\mathbf{x})+\int_{\Sigma}\int_0^{t}\,\psi^G(\tau,\mathbf{x})W(\tau,\mathbf{x},\mathbf{y})\vert\psi^F(\tau,\mathbf{y})\vert\,d\tau\,d^2\mathbf{y}-\psi^B(t,\mathbf{x})\Big\vert\\
	&\leq\psi^F(0,\mathbf{x})+\int_{\Sigma}\int_0^{t}\,\psi^G(\tau,\mathbf{x})W(\tau,\mathbf{x},\mathbf{y})\vert\psi^F(\tau,\mathbf{y})\vert\,d\tau\,d^2\mathbf{y}+\Big\vert\psi^B(t,\mathbf{x})\Big\vert\\
	&\leq\psi^F(0,\mathbf{x})+\psi^G(0,\mathbf{x})+\vert\psi^B(t,\mathbf{x})\vert\leq\psi^F(0,\mathbf{x})+\psi^G(0,\mathbf{x})+M\leq2M,
\end{align*}

\noindent
where, the inequality $\vert\int_0^T\alpha(\tau)\,\psi^F(\tau,\mathbf{x})\,d\tau\vert\leq M$ has been used.

\noindent
We have proved that $\hat{F}_TS_{T}\subset S_{T}$. Now we show that $\hat{F}_T$ is a contraction. Since, $\psi^F_1(0,\mathbf{x})=\psi^F_2(0,\mathbf{x})$ and $\psi^G_1(0,\mathbf{x})=\psi^G_2(0,\mathbf{x})$, we have

\begin{align*}
	&\quad \hat{F}_T\begin{pmatrix}
		\psi_1^F\\
		\psi_1^G\\
	\end{pmatrix}(t,\mathbf{x})-
	\hat{F}_T\begin{pmatrix}
		\psi_2^F\\
		\psi_2^G\\
	\end{pmatrix}(t,\mathbf{x})=\\
	&=\begin{pmatrix}
		-\psi^B_1(t,\mathbf{x})+\psi^B_2(t,\mathbf{x})+\int_{\Sigma}\int_0^{t}\,\Big[\psi^G_1(\tau,\mathbf{x})W(\tau,\mathbf{x},\mathbf{y})\vert\psi^F_1(\tau,\mathbf{y})\vert
		-\psi^G_2(\tau,\mathbf{x})W(\tau,\mathbf{x},\mathbf{y})\vert\psi^F_2(\tau,\mathbf{y})\vert\Big]\,d\tau\,d^2\mathbf{y}\\
		-\int_{\Sigma}\int_0^{t}\,\Big[\psi^G_1(\tau,\mathbf{x})W(\tau,\mathbf{x},\mathbf{y})\vert\psi^F_1(\tau,\mathbf{y})\vert-\psi^G_2(\tau,\mathbf{x})W(\tau,\mathbf{x},\mathbf{y})\vert\psi^F_2(\tau,\mathbf{y})\vert\Big]\,d\tau\,d^2\mathbf{y}\\
	\end{pmatrix}.
\end{align*} 

\noindent
Notice that 

\begin{align*}
	&\Big\vert\int_{\Sigma}\int_0^{t}\,\Big[\psi^G_1(\tau,\mathbf{x})W(\tau,\mathbf{x},\mathbf{y})\vert\psi^F_1(\tau,\mathbf{y})\vert-\psi^G_2(\tau,\mathbf{x})W(\tau,\mathbf{x},\mathbf{y})\vert\psi^F_2(\tau,\mathbf{y})\vert\Big]\,d\tau\,d^2\mathbf{y}\Big\vert\\
	&\leq\int_{\Sigma}\int_0^{t}\,\Big\vert\psi^G_1(\tau,\mathbf{x})-\psi^G_2(\tau,\mathbf{x})\Big\vert\vert\psi^F_2(\tau,\mathbf{y})\vert W(\tau,\mathbf{x},\mathbf{y})\,d\tau\,d^2\mathbf{y}+\\
	&\qquad\qquad\qquad\qquad+\int_{\Sigma}\int_0^{t}\,\Big\vert\vert\psi^F_1(\tau,\mathbf{y})\vert-\vert\psi^F_2(\tau,\mathbf{y})\vert\Big\vert\psi^G_1(\tau,\mathbf{x})W(\tau,\mathbf{x},\mathbf{y})\,d\tau\,d^2\mathbf{y}\\
	&\leq\|\psi^G_1-\psi^G_2\|_\infty 2M\Sigma\bar{W}T+\|\psi^F_1-\psi^F_2\|_\infty M\Sigma\bar{W}T\le \frac{1}{3}\|\psi^G_1-\psi^G_2\|_\infty+\frac{1}{6}\|\psi^F_1-\psi^F_2\|_\infty.
\end{align*}

\noindent
Therefore, 

\begin{align*}
	\|\widetilde{\psi}^F_1-\widetilde{\psi}^F_2\|_\infty&\leq\int_{0}^t\alpha(\tau,\mathbf{x})\vert\,\psi^F_1(\tau,\mathbf{x})-\psi^F_2(\tau,\mathbf{x})\vert\,d\tau+\frac{1}{3}\|\psi^G_1-\psi^G_2\|_\infty+\frac{1}{6}\|\psi^F_1-\psi^F_2\|_\infty\\
	&\le\frac{1}{2}\|\,\psi^F_1-\psi^F_2\|_\infty+\frac{1}{4}\|\psi^G_1-\psi^G_2\|_\infty+\frac{1}{6}\|\psi^F_1-\psi^F_2\|_\infty\\
	&<\max\{\|\psi^F_1-\psi^F_2\|_\infty,\|\psi^G_1-\psi^G_2\|_\infty\}=\|(\psi^F_1,\psi^G_1)-(\psi^F_1,\psi^G_2)\|_\infty.
\end{align*}

\noindent
Analogously,

\begin{align*}
	\|\widetilde{\psi}^G_1-\widetilde{\psi}^G_2\|_\infty&\leq\frac{1}{3}\|\psi^G_1-\psi^G_2\|_\infty+\frac{1}{6}\|\psi^F_1-\psi^F_2\|_\infty<\|(\psi^F_1,\psi^G_1)
	-(\psi^F_1,\psi^G_2)\|_\infty.
\end{align*}

\noindent
Therefore, $\hat{F}_T$ is a contraction since

\begin{align*} 
	\|(\widetilde{\psi}^F_1,\widetilde{\psi}^G_1)-(\widetilde{\psi}^F_2,\widetilde{\psi}^G_2)\|_\infty&=\max\{\|\widetilde{\psi}^F_1-\widetilde{\psi}^F_2\|_\infty,\|\widetilde{\psi}^G_1-\widetilde{\psi}^G_2\|_\infty\}<\|(\psi^F_1,\psi^G_1)-(\psi^F_2,\psi^G_2)\|_\infty.
\end{align*}

\noindent 
By Banach's fixed point theorem, there is a unique fixed point $\big(\psi_0^F(t,\mathbf{x}),\psi_0^G(t,\mathbf{x})\big)$, $t\in[0,T]$, for the operator $\hat{F}_T$. Notice that $\psi_0^F(t,\mathbf{x})\geq 0$ for every $t\in[0,T]$ (see the next subsection) and therefore, see (\ref{S1}), $\psi_0^F(t,\mathbf{x})+\psi_0^G(t,\mathbf{x})\leq\psi^F(0,\mathbf{x})+\psi^G(0,\mathbf{x})\leq M$ for all $t\in[0,T]$. Now, we can repeat the reasoning with $f(\mathbf{x})=\psi_0^F(T,\mathbf{x})$, $g(\mathbf{x})=\psi_0^G(T,\mathbf{x})$ as the initial condition and 
$\hat{F}_{2T}:L^\infty([T,2T]\times\Sigma,dt\,d^2\mathbf{x})\oplus L^\infty([T,2T]\times\Sigma,dt\,d^2\mathbf{x})\to L^\infty([T,2T]\times\Sigma,dt\,d^2\mathbf{x})\oplus L^\infty([T,2T]\times\Sigma,dt\,d^2\mathbf{x})$  defined by

\begin{align}\label{S2}
	\begin{pmatrix}
		\widetilde{\psi}^F(t,\mathbf{x})\\
		\widetilde{\psi}^G(t,\mathbf{x})\\
	\end{pmatrix}&:=\hat{F}_{2T}\begin{pmatrix}
		\psi^F\\
		\psi^G\\
	\end{pmatrix}(t,\mathbf{x})=\notag\\
	&=\begin{pmatrix}
		\psi_0^F(T,\mathbf{x})-\int_T^{t}\alpha(\tau,\mathbf{x})\psi^F(\tau,\mathbf{x})\,d\tau+\int_{\Sigma}\int_T^{t}\,\psi^G(\tau,\mathbf{x})W(\tau,\mathbf{x},\mathbf{y})
		\vert\psi^F(\tau,\mathbf{y})\vert\,d\tau\,d^2\mathbf{y}\\
		\psi_0^G(T,\mathbf{x})-\int_{\Sigma}\int_T^{t}\,\psi^G(\tau,\mathbf{x})W(\tau,\mathbf{x},\mathbf{y})\vert\psi^F(\tau,\mathbf{y})\vert\,d\tau\,d^2\mathbf{y}\\
	\end{pmatrix}
\end{align}

\noindent
and such that $\hat{F}_{2T}S_{2T}\subset S_{2T}$.  We obtain a fixed point $\big(\psi_1^F(t,\mathbf{x}),\psi_1^G(t,\mathbf{x})\big)$, $t\in[T,2T]$, for the operator $\hat{F}_{2T}$ such that $\big(\psi_1^F(T,\mathbf{x}\big),\psi_1^G(T,\mathbf{x})\big)=\big(\psi_0^F(T,\mathbf{x}),\psi_0^G(T,\mathbf{x})\big)$. The procedure can be iterated and at step $n+1$ we obtain a fixed point  $\big(\psi_{n}^F(t,\mathbf{x}),\psi_{n}^G(t,\mathbf{x})\big)$, $t\in[nT, (n+1)T]$, for the operator $\hat{F}_{(n+1)T}$ such that $\big(\psi_{n}^F(nT,\mathbf{x}\big),\psi_{n}^G(nT,\mathbf{x})\big)=\big(\psi_{n-1}^F(nT,\mathbf{x}),\psi_{n-1}^G(nT,\mathbf{x})\big)$, and

\begin{equation}\label{S3}
	\begin{pmatrix}
		\psi_n^F(t,\mathbf{x})\\
		\psi_n^G(t,\mathbf{x})\\
	\end{pmatrix}=
	\begin{pmatrix}
		\psi_{n-1}^F(nT,\mathbf{x})-\int_{nT}^{t}\alpha(\tau,\mathbf{x})\psi_n^F(\tau,\mathbf{x})\,d\tau+\int_{\Sigma}\int_{nT}^{t}\,\psi_n^G(\tau,\mathbf{x})W(\tau,\mathbf{x},\mathbf{y})\vert\psi_n^F(\tau,\mathbf y)\vert\,d\tau\,d^2\mathbf{y}\\
		\psi_{n-1}^G(nT,\mathbf{x})-\int_{\Sigma}\int_{nT}^{t}\,\psi_n^G(\tau,\mathbf{x})W(\tau,\mathbf{x},\mathbf{y})\vert\psi_n^F(\tau,\mathbf y)\vert\,d\tau\,d^2\mathbf{y}
	\end{pmatrix}.
\end{equation}
\noindent

It is straightforward to show that the vector $\big(\bar{\psi}^F(t,\mathbf{x}),\bar{\psi}^G(t,\mathbf{x})\big)=\big(\psi^F_k(t,\mathbf{x}),\psi^G_k(t,\mathbf{x})\big)$, $t\in[kT,(k+1)T]$, $k\in\mathbb{N}$, is the unique fixed point for the operator $\hat{F}$. First, by induction, we prove that $\hat{F}\big(\bar{\psi}^F(t,\mathbf{x}),\bar{\psi}^G(t,\mathbf{x})\big)=\big(\bar{\psi}^F(t,\mathbf{x}),\bar{\psi}^G(t,\mathbf{x})\big)$. At step 1, $k=0$, for every $t\in[0,T]$,

\begin{align*}
	\begin{pmatrix}
		\bar\psi^F(t,\mathbf{x})\\
		\bar\psi^G(t,\mathbf{x})\\
	\end{pmatrix}=	
	\begin{pmatrix}
		\psi_0^F(t,\mathbf{x})\\
		\psi_0^G(t,\mathbf{x})\\
	\end{pmatrix}&=
	\begin{pmatrix}
		\psi^F_0(0, \mathbf {x})-\int_{0}^{t}\alpha(\tau)\psi_0^F(\tau,\mathbf{x})\,d\tau+\int_{\Sigma}\int_{0}^{t}\,\psi_0^G(\tau,\mathbf{x})W(\tau,\mathbf{x},\mathbf{y})\vert\psi_0^F(\tau,\mathbf y)\vert\,d\tau\,d^2\mathbf{x}\\
		\psi_{0}^G(0,\mathbf{x})-\int_{\Sigma}\int_0^{t}\,\psi_0^G(\tau,\mathbf{x})W(\tau,\mathbf{x},\mathbf{y})\vert\psi_0^F(\tau,\mathbf y)\vert\,d\tau\,d^2\mathbf{y}\\
	\end{pmatrix}\\
	&=\begin{pmatrix}
		\bar{\psi}^F(0, \mathbf x)-\int_{0}^{t}\alpha(\tau)\bar{\psi}^F(\tau,\mathbf{x})\,d\tau+\int_{\Sigma}\int_{0}^{t}\,\bar{\psi}^G(\tau,\mathbf{x})W(\tau,\mathbf{x},\mathbf{y})\vert\bar{\psi}^F(\tau,\mathbf y)\vert\,d\tau\,d^2\mathbf{x}\\
		\bar{\psi}^G(0,\mathbf{x})-\int_{\Sigma}\int_0^{t}\,\bar{\psi}^G(\tau,\mathbf{x})W(\tau,\mathbf{x},\mathbf{y})\vert\bar{\psi}^F(\tau,\mathbf y)\vert\,d\tau\,d^2\mathbf{y}.
	\end{pmatrix}
\end{align*}

\noindent
Inductive step: Suppose that for every $t\in[nT,(n+1)T]$,

\begin{align*}
	\begin{pmatrix}
		\bar\psi^F(t,\mathbf{x})\\
		\bar\psi^G(t,\mathbf{x})\\
	\end{pmatrix}&=
	\begin{pmatrix}
		\psi_n^F(t,\mathbf{x})\\
		\psi_n^G(t,\mathbf{x})\\
	\end{pmatrix}\\
	&=\begin{pmatrix}
		\psi^F_{n}(nT, \mathbf x)-\int_{nT}^{t}\alpha(\tau)\psi_n^F(\tau,\mathbf{x})\,d\tau+\int_{\Sigma}\int_{nT}^{t}\,\psi_n^G(\tau,\mathbf{x})W(\tau,\mathbf{x},\mathbf{y})\vert\psi_n^F(\tau,\mathbf y)\vert\,d\tau\,d^2\mathbf{x}\\
		\psi_{n}^G(nT,\mathbf{x})-\int_{\Sigma}\int_{nT}^{t}\,\psi_n^G(\tau,\mathbf{x})W(\tau,\mathbf{x},\mathbf{y})\vert\psi_n^F(\tau,\mathbf y)\vert\,d\tau\,d^2\mathbf{y}\\
	\end{pmatrix}\\
	&=\begin{pmatrix}
		\bar{\psi}^F(0, \mathbf x)-\int_{0}^{t}\alpha(\tau)\bar{\psi}^F(\tau,\mathbf{x})\,d\tau+\int_{\Sigma}\int_{0}^{t}\,\bar{\psi}^G(\tau,\mathbf{x})W(\tau,\mathbf{x},\mathbf{y})\vert\bar{\psi}^F(\tau,\mathbf y)\vert\,d\tau\,d^2\mathbf{x}\\
		\bar{\psi}^G(0,\mathbf{x})-\int_{\Sigma}\int_0^{t}\,\bar{\psi}^G(\tau,\mathbf{x})W(\tau,\mathbf{x},\mathbf{y})\vert\bar{\psi}^F(\tau,\mathbf y)\vert\,d\tau\,d^2\mathbf{y}\\
	\end{pmatrix}.
\end{align*}

\noindent
Then,

\begin{align*}
	\nonumber
	&\begin{pmatrix}
		\bar\psi^F((n+1)T,\mathbf{x})\\
		\bar\psi^G((n+1)T,\mathbf{x})\\
	\end{pmatrix}=\begin{pmatrix}
		\psi_n^F((n+1)T,\mathbf{x})\\
		\psi_n^G((n+1)T,\mathbf{x})\\
	\end{pmatrix}=\\
	&=\begin{pmatrix}
		\bar{\psi}^F(0, \mathbf x)-\int_{0}^{(n+1)T}\alpha(\tau)\bar{\psi}^F(\tau,\mathbf{x})\,d\tau+\int_{\Sigma}\int_{0}^{(n+1)T}\,\bar{\psi}^G(\tau,\mathbf{x})W(\tau,\mathbf{x},\mathbf{y})\vert\bar{\psi}^F(\tau,\mathbf y)\vert\,d\tau\,d^2\mathbf{x}\\
		\bar{\psi}^G(0,\mathbf{x})-\int_{\Sigma}\int_0^{(n+1)T}\,\bar{\psi}^G(\tau,\mathbf{x})W(\tau,\mathbf{x},\mathbf{y})\vert\bar{\psi}^F(\tau,\mathbf y)\vert\,d\tau\,d^2\mathbf{y}
	\end{pmatrix}.
\end{align*}

\noindent
Hence, for every $t\in[(n+1)T,(n+2)T]$,

\begin{align*}
	&\begin{pmatrix}
		\bar{\psi}^F(0,x)-\int_{0}^{t}\alpha(\tau)\bar{\psi}^F(\tau,\mathbf{x})\,d\tau+\int_{\Sigma}\int_{0}^{t}\,\bar{\psi}^G(\tau,\mathbf{x})W(\tau,\mathbf{x},\mathbf{y})\vert\bar{\psi}^F(\tau,\mathbf y)\vert\,d\tau\,d^2\mathbf{x}\\
		\bar{\psi}^G(0,\mathbf{x})-\int_{\Sigma}\int_0^{t}\,\bar{\psi}^G(\tau,\mathbf{x})W(\tau,\mathbf{x},\mathbf{y})\vert\bar{\psi}^F(\tau,\mathbf y)\vert\,d\tau\,d^2\mathbf{y}\\
	\end{pmatrix}=\\
	&=\begin{pmatrix}
		\bar{\psi}^F(0,x)-\int_{0}^{(n+1)T}\alpha(\tau)\bar{\psi}^F(\tau,\mathbf{x})\,d\tau+\int_{\Sigma}\int_{0}^{(n+1)T}\,\bar{\psi}^G(\tau,\mathbf{x})W(\tau,\mathbf{x},\mathbf{y})\vert\bar{\psi}^F(\tau,\mathbf y)\vert\,d\tau\,d^2\mathbf{x}\\
		\bar{\psi}^G(0,\mathbf{x})-\int_{\Sigma}\int_0^{(n+1)T}\,\bar{\psi}^G(\tau,\mathbf{x})W(\tau,\mathbf{x},\mathbf{y})\vert\bar{\psi}^F(\tau,\mathbf y)\vert\,d\tau\,d^2\mathbf{y}\\
	\end{pmatrix}+\\
	&\quad\quad\quad\quad+\begin{pmatrix}
		-\int_{(n+1)T}^t\alpha(\tau)\bar{\psi}^F(\tau,\mathbf{x})\,d\tau+\int_{\Sigma}\int_{(n+1)T}^t\,\bar{\psi}^G(\tau,\mathbf{x})W(\tau,\mathbf{x},\mathbf{y})\vert\bar{\psi}^F(\tau,\mathbf y)\vert\,d\tau\,d^2\mathbf{x}\\
		-\int_{\Sigma}\int_{(n+1)T}^t\,\bar{\psi}^G(\tau,\mathbf{x})W(\tau,\mathbf{x},\mathbf{y})\vert\bar{\psi}^F(\tau,\mathbf y)\vert\,d\tau\,d^2\mathbf{y}\\
	\end{pmatrix}\\
	&=\begin{pmatrix}
		\psi^F_{n}((n+1)T,x)-\int_{(n+1)T}^{t}\alpha(\tau)\psi_n^F(\tau,\mathbf{x})\,d\tau+\int_{\Sigma}\int_{(n+1)T}^{t}\,\psi_n^G(\tau,\mathbf{x})W(\tau,\mathbf{x},\mathbf{y})\vert\psi_n^F(\tau,\mathbf y)\vert\,d\tau\,d^2\mathbf{x}\\
		\psi_{n}^G((n+1)T,\mathbf{x})-\int_{\Sigma}\int_{(n+1)T}^{t}\,\psi_n^G(\tau,\mathbf{x})W(\tau,\mathbf{x},\mathbf{y})\vert\psi_n^F(\tau,\mathbf y)\vert\,d\tau\,d^2\mathbf{y}\\
	\end{pmatrix}\\
	&=\begin{pmatrix}
		\psi_{n+1}^F(t,\mathbf{x})\\
		\psi_{n+1}^G(t,\mathbf{x})
	\end{pmatrix}.
\end{align*}

\noindent
It remains to prove the uniqueness. If $\big(\hat{\psi}^F(t,\mathbf{x}),\hat{\psi}^G(t,\mathbf{x})\big)$ is a second fixed point for the operator $\hat{F}$, then, for every $n\in\mathbb{N}$, its restriction $\big(\hat{\psi}^F(t,\mathbf{x}),\hat{\psi}^G(t,\mathbf{x})\big)_{[nT,(n+1)T]}$ to the interval $[nT,(n+1)T]$ is a fixed point for $\hat{F}_{(n+1)T}$. Indeed, for every $t\in[nT,(n+1)T]$,

\begin{align*}
	\hat{F}_{(n+1)T}\begin{pmatrix}
		\hat{\psi}^F(t,\mathbf{x})\\
		\hat{\psi}^G(t,\mathbf{x})
	\end{pmatrix}&=\begin{pmatrix}
		\hat{\psi}^F(nT,x)-\int_{nT}^{t}\alpha(\tau)\hat{\psi}^F(\tau,\mathbf{x})\,d\tau+\int_{\Sigma}\int_{nT}^{t}\,\hat{\psi}^G(\tau,\mathbf{x})W(\tau,\mathbf{x},\mathbf{y})\vert\hat{\psi}^F(\tau,\mathbf y)\vert\,d\tau\,d^2\mathbf{x}\\
		\hat{\psi}^G(nT,\mathbf{x})-\int_{\Sigma}\int_{nT}^{t}\,\hat{\psi}^G(\tau,\mathbf{x})W(\tau,\mathbf{x},\mathbf{y})\vert\hat{\psi}^F(\tau,\mathbf y)\vert\,d\tau\,d^2\mathbf{y}
	\end{pmatrix}=\\
	&=\begin{pmatrix}
		\hat{\psi}^F(0,x)-\int_{0}^{nT}\alpha(\tau)\hat{\psi}^F(\tau,\mathbf{x})\,d\tau+\int_{\Sigma}\int_{0}^{nT}\,\hat{\psi}^G(\tau,\mathbf{x})W(\tau,\mathbf{x},\mathbf{y})\vert\hat{\psi}^F(\tau,\mathbf y)\vert\,d\tau\,d^2\mathbf{x}\\
		\hat{\psi}^G(0,\mathbf{x})-\int_{\Sigma}\int_0^{nT}\,\bar{\psi}^G(\tau,\mathbf{x})W(\tau,\mathbf{x},\mathbf{y})\vert\bar{\psi}^F(\tau,\mathbf y)\vert\,d\tau\,d^2\mathbf{y}\\
	\end{pmatrix}+\\
	&\qquad\qquad+\begin{pmatrix}
		-\int_{nT}^{t}\alpha(\tau)\hat{\psi}^F(\tau,\mathbf{x})\,d\tau+\int_{\Sigma}\int_{nT}^{t}\,\hat{\psi}^G(\tau,\mathbf{x})W(\tau,\mathbf{x},\mathbf{y})\vert\hat{\psi}^F(\tau,\mathbf y)\vert\,d\tau\,d^2\mathbf{x}\\
		-\int_{\Sigma}\int_{nT}^{t}\,\hat{\psi}^G(\tau,\mathbf{x})W(\tau,\mathbf{x},\mathbf{y})\vert\hat{\psi}^F(\tau,\mathbf y)\vert\,d\tau\,d^2\mathbf{y}
	\end{pmatrix}\\
	&=\begin{pmatrix}
		\hat{\psi}^F(0,x)-\int_{0}^{t}\alpha(\tau)\hat{\psi}^F(\tau,\mathbf{x})\,d\tau+\int_{\Sigma}\int_{0}^{t}\,\hat{\psi}^G(\tau,\mathbf{x})W(\tau,\mathbf{x},\mathbf{y})\vert\hat{\psi}^F(\tau,\mathbf y)\vert\,d\tau\,d^2\mathbf{x}\\
		\hat{\psi}^G(0,\mathbf{x})-\int_{\Sigma}\int_{0}^{t}\,\hat{\psi}^G(\tau,\mathbf{x})W(\tau,\mathbf{x},\mathbf{y})\vert\hat{\psi}^F(\tau,\mathbf y)\vert\,d\tau\,d^2\mathbf{y}
	\end{pmatrix}\\
	&=\hat{F}\begin{pmatrix}
		\hat{\psi}^F(t,\mathbf{x})\\
		\hat{\psi}^G(t,\mathbf{x})
	\end{pmatrix}=\begin{pmatrix}
		\hat{\psi}^F(t,\mathbf{x})\\
		\hat{\psi}^G(t,\mathbf{x})
	\end{pmatrix}.
\end{align*}

\noindent 
Since $\hat{F}_{(n+1)T}$ admits a unique fixed point, it must be $\big(\hat{\psi}^F(t,\mathbf{x}),\hat{\psi}^G(t,\mathbf{x})\big)_{[nT,(n+1)T]}=\big(\bar\psi_n^F(t,\mathbf{x}),\bar\psi_n^G(t,\mathbf{x})\big)$.

\subsubsection*{Positive definitness of the solution}\label{positive}
Here, we show that the solution $\psi^F(t,\mathbf{x})$, whose existence has been proved above, is positive definite. 

Suppose there is a point $(t,\mathbf{x})$ such that $\psi^F(t,\mathbf{x})<0$. Then, by the continuity of the functions and by $\psi^F(0,\mathbf{x})\geq0$, there should be $t_0>0$ and $h>0$, such that $\psi^F(t_0,\mathbf{x})=0$ and $\psi^F(t,\mathbf{x})<0$, $\forall t\in(t_0,t_0+h]$. Then,

\begin{equation}\label{a}
	0=\psi^F(t_0,\mathbf{x})=\psi^F(0,\mathbf{x})-\int_0^{t_0}\alpha(\tau)\psi^F(\tau,\mathbf{x})\,d\tau+\int_{\Sigma}\int_0^{t_0}\,\psi^G(\tau,\mathbf{x})W(\tau,\mathbf{x},\mathbf{y})
	|\psi^F(\tau,\mathbf y)|\,d\tau\,d^2\mathbf{y}.
\end{equation}

\noindent
Moreover, for every $\epsilon\in(0,h]$, 

\begin{align*}
	0>\psi^F(t_0+\epsilon,\mathbf{x})&=\psi^F(0,\mathbf{x})-\int_0^{t_0+\epsilon}\alpha(\tau)\psi^F(\tau,\mathbf{x})\,d\tau+\int_{\Sigma}\int_0^{t_0+\epsilon}\,\psi^G(\tau,\mathbf{x})W(\tau,\mathbf{x},\mathbf{y})\vert\psi^F(\tau,\mathbf y)\vert\,d\tau\,d^2\mathbf{y}\\
	&=\psi^F(0,\mathbf{x})-\int_0^{t_0}\alpha(\tau)\psi^F(\tau,\mathbf{x})\,d\tau+\int_{\Sigma}\int_0^{t_0}\,\psi^G(\tau,\mathbf{x})W(\tau,\mathbf{x},\mathbf{y})\vert\psi^F(\tau,\mathbf y)\vert\,d\tau\,d^2\mathbf{y}\\
	&-\int_{t_0}^{t_0+\epsilon}\alpha(\tau)\psi^F(\tau,\mathbf{x})\,d\tau+\int_{\Sigma}\int_{t_0}^{t_0+\epsilon}\,\psi^G(\tau,\mathbf{x})W(\tau,\mathbf{x},\mathbf{y})\vert\psi^F(\tau,\mathbf y)\vert\,d\tau\,d^2\mathbf{y}\\
	&=-\int_{t_0}^{t_0+\epsilon}\alpha(\tau)\psi^F(\tau,\mathbf{x})\,d\tau+\int_{\Sigma}\int_{t_0}^{t_0+\epsilon}\,\psi^G(\tau,\mathbf{x})W(\tau,\mathbf{x},\mathbf{y})\vert\psi^F(\tau,\mathbf y)\vert\,d\tau\,d^2\mathbf{y}\\
	&=-\epsilon\alpha(\hat{t})\psi^F(\hat{t},\mathbf{x})+\int_{\Sigma}\int_{t_0}^{t_0+\epsilon}\,\psi^G(\tau,\mathbf{x})W(\tau,\mathbf{x},\mathbf{y})\vert\psi^F(\tau,\mathbf y)\vert\,d\tau\,d^2\mathbf{y},
\end{align*}

\noindent
where, $\hat{t}\in(t_0,t_0+\epsilon]$. Since $\psi^F(\hat{t},\mathbf{x})<0$ and $\alpha(t)\geq 0$, it must be

$$0>\psi^F(t+\epsilon,\mathbf{x}_0)=-\epsilon\alpha(\hat{t})\psi^F(\hat{t},\mathbf{x})+\int_{\Sigma}\int_{t_0}^{t_0+\epsilon}\,\psi^G(\tau,\mathbf{x})W(\tau,\mathbf{x},\mathbf{y})\vert\psi^F(\tau,\mathbf y)\vert\,d\tau\,d^2\mathbf{y}>0,$$

\noindent
which is an absurd. As a consequence, $\psi^F(t,\mathbf{x})\geq0$ for every $\mathbf{x}\in\Sigma$ and $t\in\mathbb{R}_+$. 

\subsection*{Existence and uniqueness of solutions II}

Let us assume $0\leq W(t,\mathbf{x},\mathbf{y})\leq\bar{W}\in\mathbb{R}_+$. 
In order to prove that system of equations (\ref{evolutionFG}) admits a unique solution, let us assign a non-negative and bounded function $M:\Sigma\rightarrow \mathbb R^+$, with $\overline M:={\rm sup}_{\mathbf x\in \Sigma}\{M(\mathbf x)\}$,
and consider the set $S$ of functions $\psi^F, \psi^G$ satisfying the following conditions:
\begin{itemize}
	\item $\psi^F, \psi^G\in L^\infty\big(\mathbb R\times \Sigma , dt\,d^2\mathbf x\big)$, continuous, with $\psi^G(t,\mathbf x)$ nonincreasing with respect to $t$, $\forall \mathbf x\in \Sigma,$ 
	\item $0\le\psi^F(t,\mathbf x), \psi^G(t,\mathbf x)\le M(\mathbf x)$, $\forall t\in \mathbb R,\,\,\forall \mathbf x\in \Sigma,$
	\item $\psi^F(0,\mathbf x)+\psi^G(0,\mathbf x)= M(\mathbf x),$ $\forall \mathbf x\in \Sigma$.
\end{itemize}
Let us define $\mathcal L^\infty:=L^\infty\big(\mathbb R\times \Sigma , dt\,d^2\mathbf x\big)\oplus L^\infty\big(\mathbb R\times \Sigma , dt\,d^2\mathbf x\big)$, with the norm
$$\|(\psi^F,\psi^G)\|_\infty=\max\{\|\psi^F\|_\infty,\|\psi^G\|_\infty\}.$$
Note that $\int_{\Sigma}\int_0^{t}\psi^G(\tau,\mathbf{x})\,W(\tau,\mathbf{x},\mathbf{y})\psi^F(\tau,\mathbf{y})\,d\tau\,d^2\mathbf{y}$ is a nondecreasing function of $t$ and introduce the function $T(\mathbf x),\,\,\mathbf x\in\Sigma$, which at $ \mathbf x$ is equal to the time $t$ at which
$$\int_{\Sigma}\int_0^{t}\psi^G(\tau,\mathbf{x})\,W(\tau,\mathbf{x},\mathbf{y})\psi^F(\tau,\mathbf{y})\,d\tau\,d^2\mathbf{y}=\psi^G(0,\mathbf{x}),$$
if such a time exists, equal to $\infty$ otherwise.\\
Define also the function
$$
\eta(t,\mathbf x)=\Bigg\{
\begin{array}{l}
	1\,\quad{\rm if} \quad t\le T(\mathbf x),\\
	0\quad	{\rm otherwise},
\end{array}	
$$	
and the operators
\begin{align*}
	(\psi^F(t,\mathbf x),\psi^G(t,\mathbf x))& \mapsto R_G(t,\mathbf x)=\int_{\Sigma}\int_0^{t}\psi^G(\tau,\mathbf{x})\,W(\tau,\mathbf{x},\mathbf{y})\psi^F(\tau,\mathbf{y})\eta(\tau,\mathbf{x})\,d\tau\,d^2\mathbf{y},\\
	(\psi^F(t,\mathbf x),\psi^G(t,\mathbf x))& \mapsto R_F(t,\mathbf x)=\int_{\Sigma}\int_0^{t}\psi^G(\tau,\mathbf{x})\,W(\tau,\mathbf{x},\mathbf{y})\psi^F(\tau,\mathbf{y})\pi(t-\tau)\eta(\tau,\mathbf{x})\,d\tau\,d^2\mathbf{y},\\
	\mathcal T:(\psi^F(t,\mathbf x),\psi^G(t,\mathbf x))&\mapsto	(\widetilde\psi^F(t,\mathbf x),\widetilde\psi^G(t,\mathbf x))=(\psi^F(0,\mathbf x)\pi(t)+R_F(t,\mathbf x),\psi^G(0,\mathbf x)-R_G(t,\mathbf x)).
\end{align*}
Note that if $T(\mathbf x)$ is finite at $\mathbf x$, then $\widetilde\psi^G(t,\mathbf x)=0$ for $t\ge T(\mathbf x)$. It is easy to show that 
$(\psi^F(t,\mathbf x),\psi^G(t,\mathbf x))\in S$ is a solution of system \eqref{evolutionFG} if and only it is a fixed point of the operator $\mathcal T.$ Moreover,
the following properties hold.
\begin{property}
	$\widetilde\psi^G$ is a nonincreasing function of $t$, and $0\le\widetilde\psi^G(t,\mathbf x)\le\psi^G(0,\mathbf x)\le M(\mathbf x)\le \overline M.$
\end{property}
{\bf Proof}. From its definition, it follows that $R_G(t,\mathbf x)\le \psi^G(0,\mathbf x)$. 
\begin{property}
	$0\le\widetilde\psi^F(t,\mathbf x)\le M(\mathbf x)\le \overline M.$
\end{property}
{\bf Proof}. $\widetilde\psi^F(t,\mathbf x)\le \psi^F(0,\mathbf x) +R_G(t,\mathbf x)\le \psi^F(0,\mathbf x) +\psi^G(0,\mathbf x)= M(\mathbf x)\le \overline M.$
\mbox{ } \\[0.3cm]
From the previous properties it follows that $\mathcal T S\subseteq S.$ Now let us take $\overline t>0$ to be suitably chosen and consider $0\le t\le \overline t,$ one has: 
\begin{eqnarray*}
	&\vert\widetilde\psi_2^F(t,\mathbf x)-\widetilde\psi_1^F(t,\mathbf x)\vert
	=\vert\int_{\Sigma}\int_0^{t}\,W(\tau,\mathbf{x},\mathbf{y})\Big[\psi^G_2(\tau,\mathbf{x})\psi^F_2(\tau,\mathbf{y})
	-\psi^G_1(\tau,\mathbf{x})\psi^F_1(\tau,\mathbf{y})\Big]\eta(\tau,\mathbf x)\pi(t-\tau)\,d\tau\,d^2\mathbf{y}	\vert\\
	& \le\vert\int_{\Sigma}\int_0^{t}\,W(\tau,\mathbf{x},\mathbf{y})\Big[\psi^G_2(\tau,\mathbf{x})\psi^F_2(\tau,\mathbf{y})
	-\psi^G_1(\tau,\mathbf{x})\psi^F_1(\tau,\mathbf{y})\Big]\,d\tau\,d^2\mathbf{y}	\vert\ge \vert\widetilde\psi_2^G(t,\mathbf x)-\widetilde\psi_1^G(t,\mathbf x)\vert.
\end{eqnarray*}
Being
\begin{align*}
	&\Big\vert\int_{\Sigma}\int_0^{t}\,W(\tau,\mathbf{x},\mathbf{y})\Big[\psi^G_2(\tau,\mathbf{x})\psi^F_2(\tau,\mathbf{y})
	-\psi^G_1(\tau,\mathbf{x})\psi^F_1(\tau,\mathbf{y})\Big]\,d\tau\,d^2\mathbf{y}\Big\vert\\
	&\le \int_{\Sigma}\int_0^{t}\,W(\tau,\mathbf{x},\mathbf{y})\Big[\vert\psi^G_2(\tau,\mathbf{x})\psi^F_2(\tau,\mathbf{y})-
	\psi^G_2(\tau,\mathbf{x})\psi^F_1(\tau,\mathbf{y})\vert
	+\vert\psi^G_2(\tau,\mathbf{x})\psi^F_1(\tau,\mathbf{y})-\psi^G_1(\tau,\mathbf{x})\psi^F_1(\tau,\mathbf{y})\vert\Big]\,d\tau\,d^2\mathbf{y}\\
	&\le 2 \overline M \|(\psi^F_2-\psi^F_1,\psi^G_2-\psi^G_1)\|_\infty \int_{\Sigma}\int_0^{t}\,W(\tau,\mathbf{x},\mathbf{y})\,d\tau\,d^2\mathbf{y} \le 2 \overline M\vert \Sigma\vert \overline W \,\overline t	\|(\psi^F_2-\psi^F_1,\psi^G_2-\psi^G_1)\|_\infty,
\end{align*}
Taking $\bar t<\frac{1}{2 \overline M\vert \Sigma\vert \overline W}$, $\mathcal T$ is a contraction and the solution exists and is unique up to time $\overline t$.
Now considering the time interval $\Big[\overline t, 2\overline t\Big]$, using the value of the solution at time $\overline t$ as initial datum,  with the same reasoning as 
in the previous proof of existence and uniqueness, one can show that the solution exists up to time $2 \overline t$, and iterating the procedure the global existence in time follows. Therefore we have proved the following 
\begin{theorem}
	For any initial datum belonging to $S$, system \eqref{evolutionFG} admits a unique solution which is global in time. 
\end{theorem} 
\subsection*{Asymptotic behaviour}
Now, let us suppose that $\pi(t)=\exp{(-\beta t)}$, with $\beta \in \mathbb R^+$, then the following holds

\begin{theorem}
	For $t$ going to infinity, the solution $\psi^F(t,\mathbf x)$ of the system of equations \eqref{evolutionFG} decays to zero, for any solution of the system. 
\end{theorem} 
{\bf Proof.} One has 
\begin{align*}
		\frac{\partial \psi^F }{\partial t}&=-\beta \psi^F_0\pi(t)+
	\int_{\Sigma}\,\psi^G(t,\mathbf{x})W(t,\mathbf{x},\mathbf{y})\psi^F(t,\mathbf{y})d^2\mathbf{y}
	-\beta \int_{\Sigma}\int_0^{t}\,\psi^G(\tau,\mathbf{x})W(\tau,\mathbf{x},\mathbf{y})\psi^F(\tau,\mathbf{y})
	\pi(t-\tau)\,d\tau\,d^2\mathbf{y}\\
	&=-\beta\psi^F(t,\mathbf{x})-\frac{\partial \psi^G }{\partial t}(t,\mathbf{x}),
\end{align*}
from which 
\begin{eqnarray*}
	& 	\frac{\partial \psi^F }{\partial t}+\frac{\partial \psi^G }{\partial t}= -\beta\psi^F\le 0.
\end{eqnarray*}
Therefore, for any $\mathbf x \in \Sigma$, there exists $\lim_{t\rightarrow \infty}(\psi^F+\psi^G)(t,\mathbf{x})$ and is finite and nonnegative, which implies 
that $\lim_{t\rightarrow \infty}(\frac{\partial \psi^F}{\partial t}+\frac{\partial \psi^G}{\partial t})=0$, hence, from the previous equality
we can conclude that $\lim_{t\rightarrow \infty} \psi^F(t,\mathbf{x})=0,\,\,\,\forall \mathbf x \in \Sigma$.

\noindent 
By using exactly the same reasoning we obtain the equivalent result for the memoryless case.

\begin{theorem}
	For $t$ going to infinity, the solution $\psi^F(t,\mathbf x)$ of the system of \eqref{system} decays to zero, $\forall \mathbf x\in \Sigma$ and  for any solution of the system. 
\end{theorem}

\end{document}